\documentclass[prc,aps,float,twocolumn,showpacs,nofootinbib,superscriptaddress]{revtex4}
\usepackage{amsmath}
\usepackage{amssymb}
\usepackage{amsfonts}
\usepackage{graphicx}
\usepackage[usenames,dvipsnames]{color}
\usepackage[colorlinks,linkcolor=blue,anchorcolor=blue,citecolor=blue]{hyperref}
\allowdisplaybreaks

\usepackage{ulem}

\newcommand{\nuc}[2]{$^{#1}${#2}}

\newcommand{\beq}{\begin{equation}}
\newcommand{\eeq}{\end{equation}}
\newcommand{\beqn}{\begin{eqnarray}}
\newcommand{\eeqn}{\end{eqnarray}}

\begin{document}
\title{Systematics of low-lying states of even-even nuclei
       in the neutron-deficient lead region from a beyond-mean-field
       calculation}

\author{J. M. Yao}
\affiliation{Physique Nucl\'eaire Th\'eorique,
             Universit\'e Libre de Bruxelles, C.P. 229, B-1050 Bruxelles,
             Belgium}
\affiliation{School of Physical Science and Technology,
 Southwest University, Chongqing, 400715 China}

\author{M. Bender}
\affiliation{Universit\'e Bordeaux, Centre d'Etudes Nucl\'eaires de
Bordeaux Gradignan, UMR5797, F-33175 Gradignan, France}
\affiliation{CNRS/IN2P3, Centre d'Etudes Nucl\'eaires de Bordeaux
Gradignan, UMR5797, F-33175 Gradignan, France}

\author{P.-H. Heenen}
\affiliation{Physique Nucl\'eaire Th\'eorique,
             Universit\'e Libre de Bruxelles, C.P. 229, B-1050 Bruxelles,
             Belgium}

%\date{9 November 2012}

\begin{abstract}
\begin{description}

\item[Background] Nuclei located in the neutron-deficient Pb region have
a complex structure, rapidly evolving as a function of neutron and
proton numbers. The most famous example is $^{186}$Pb where the three
lowest levels are $0^+$ states, the two excited $0^+$ ones being located at low
excitation energy around 600~keV. Coexisting structures with different
properties are found in the neighboring nuclei. Many experiments have
been performed over the last few years in which in-band and out-of-band
$\gamma$ transition probabilities have been measured.

\item[Purpose] A detailed interpretation of experimental data requires the
use of a method going beyond a mean-field approach that permits to determine
spectra and transition probabilities. Such methods have already been applied
to selected isotopes in this mass region. Our aim is to provide a systematic
investigation of this mass region in order to determine how well experimental
data can be understood using a state-of-the-art method for nuclear structure.

\item[Method] The starting point of our method is a set of mean-field wave
functions generated with a constraint on the axial quadrupole moment and using
a Skyrme energy density functional. Correlations beyond the mean field are
introduced by projecting mean-field wave functions on angular-momentum and
particle number and by mixing the symmetry restored wave functions
as a function of the axial quadrupole moment.

\item[Results] A detailed comparison with the available data is performed
for energies, charge radii, spectroscopic quadrupole moments, $E0$ and $E2$
transition probabilities for the isotopic chains of neutron deficient Hg,
Pb, Po and Rn. The connection between our results and the underlying
mean field is also analyzed.

\item[Conclusions]
Qualitative agreement with the data is obtained although our results
indicate that the actual energy density functionals have to be improved further to
achieve a quantitative agreement.

\end{description}
\end{abstract}


\maketitle

%%%%%%%%%%%%%%%%%%%%%%%%%%%%%%%%%%%%%%%%%%%%%%%%%%%%%%%%%%%%%%%%%%%%%%%%%%

\section{Introduction}

The study of the low-energy spectrum of neutron-deficient nuclei around
$^{186}$Pb has unveiled a rich variety of collective levels. In particular,
several $0^+$ states coexist at low excitation energy, which are in several
cases the band head of a rotational sequence. This experimental result is
interpreted as a manifestation of ``shape coexistence''; i.e.,\ a condition where states
at similar excitation energy have distinctly different intrinsic shapes~\cite{Heyde83,Heyde88,Wood92,Julin01,Heyde11}. Many data support this
interpretation. Direct evidence is, indeed, provided by the systematics
of $\alpha$-decay fine structure~\cite{Duppen00,Andreyev00,Julin01}, by
the moments of inertia of the rotational bands built on several $0^+$
states~\cite{Julin01}, by in-band and intraband transition probabilities~\cite{Dewald03}, by $g$-factor measurements~\cite{Vyvey04,Bujor10} and by charge radii~\cite{Bonn72,Bonn76}.

Let us first recall how the low-energy spectrum of the Pb isotopes around $N=104$ is interpreted in terms of deformation.  For all isotopes, many experimental evidences indicate that the ground states are predominantly spherical~\cite{Heyde11}. The two lowest excited $0^+$ states have been
associated with two deformed structures, an oblate one lower in energy in the isotopes above $^{188}$Pb, and a prolate one that is lower in lighter Pb nuclei. The crossing of these two structures in the excitation spectrum leads to the unique situation where the three lowest levels in $^{186}$Pb are $0^+$ states~\cite{Andreyev00}.

By contrast, the ground states of the Hg isotopes down to $^{180}$Hg are interpreted as being oblate and weakly deformed with $\beta \approx -0.15$~\cite{Raman01,Grahn09}. From $^{198}$Hg down to
$^{190}$Hg,  the yrast states for a given angular momentum have almost constant excitation energies and are interpreted as the members of a rotational band based on the oblate ground state~\cite{Bea94}. This simple pattern is distorted for the lighter even isotopes through the intrusion of a strongly deformed prolate band with $\beta \approx 0.25$~\cite{Wood92,Julin01}.
The excitation energies of the states in the prolate band evolve in a nearly parabolic manner as a function of the neutron number, cf., for example, Fig.~1 in Ref.~\cite{Elseviers11} or Fig.~10 in Ref~\cite{Heyde11}.

The production rates of light Po isotopes are much smaller than those
of the corresponding Hg and Pb isotones; hence, less experimental data
have been collected, but they give a clear indication for an even more
complex evolution of their structure as a function of
neutron number. The heavier Po isotopes down to $^{196}$Po have
near-spherical ground states~\cite{Heyde11}. For $^{188-192}$Po, the
analysis of their $\alpha$-decay fine structure indicates that the
ground-state wave function contains a significant contribution from
deformed configurations~\cite{Vel03PRC}. Similar conclusions have been
drawn from charge radii that indicate an increasing softness of the
nucleus against quadrupole deformations with decreasing neutron
number~\cite{Coc11}.

Our study will be limited to even-even nuclei. Let us, however, mention
that data for odd-mass nuclei, where the low-lying spectrum can often be
interpreted by the coupling of one particle to the low-lying states of the
adjacent even-even nucleus, corroborate the interpretation of the spectral
data obtained for the light even-even isotopes of
Hg, Pb, Po and Rn~\cite{Heyde11}.

There are two different approaches to interpret the complex and
rapidly changing structure of nuclei in this mass region. One is based
on the shell model, where the emergence of low-energy intruder states
is interpreted as resulting from proton excitations across the $Z=82$
closed shell~\cite{Duppe84,Hey86,Wood92}. Although the shell model including
$n$p-$m$h excitations across the
shell gaps provides a consistent and microscopic description of the phenomenon
of shape coexistence throughout the nuclear chart~\cite{Heyde11}, the
high dimensionality of the corresponding model space renders calculations
for heavy open-shell nuclei untractable. For such nuclei, however, the
$n$p-$m$h excitations can be handled within the algebraic framework of
the interacting boson model, which truncates the shell-model space through
the approximation of \mbox{$J=0$} and \mbox{$J=2$} coupled nucleon pairs treated as
bosons~\cite{Hellemans05,Hellemans08}.

An alternative and intuitive approach is based on mean-field methods, which
provide a description of these structures in terms of shapes and deformed
shells. In this framework, the low-lying spectrum is explained by the
presence of multiple local minima in the deformation energy surface.
These minima can in turn be related to gaps in the single-particle spectrum.
Although shape coexistence can be interpreted within pure mean-field models, a
detailed description of the spectrum requires to ``go beyond the mean-field"
by means of symmetry restoration and taking into account fluctuations in
the deformation degrees of freedom. The beyond self-consistent mean-field
method that has been developed by several groups combines the projection
techniques with the generator coordinate method
(GCM)~\cite{Egido04,Bender08LH,Niksic06,Bender2008,Yao10,Rodriguez10,Bender2004}.

Throughout this work, we use the axial quadrupole moment as a collective
variable. Mean-field wave functions are constructed by Hartree-Fock
(HF)+BCS calculations. They are projected on angular-momentum and particle
number to form a basis of states that are mixed by the GCM. The interaction
used here is the SLy6 parametrization of the Skyrme interaction that
describes well the systematics of low-lying states of neutron-deficient
Pb isotopes~\cite{Duguet03,Bender04}. Selected results for some of the
nuclei discussed here have already been published earlier in
Refs.~\cite{Grahn08,Grahn09,Witte07,Rahkila10,Coc11}.

The paper is organized as follows. In Sec.~\ref{Sec.II}, we will
give a brief outline of the method used to calculate the
spectroscopic properties of low-lying states. In Sec.~\ref{Sec.III}, the
calculated deformation energy curves, low-energy excitation spectra,
charge radii, kinetic moments of inertia,
and electric monopole and quadrupole transition strengths are presented
and discussed in comparison with results of our previous study of Pb
isotopes and the available data. A summary of our findings and
conclusions is given in Sec.~\ref{Sec.IV}.

%%%%%%%%%%%%%%%%%%%%%%%%%%%%%%%%%%%%%%%%%%%%%%%%%%%%%%%%%%%%%%%%%%%%%%%%%%

\section{The method}
\label{Sec.II}

As in our previous works~\cite{Bender04}, the starting point of our method
is a set of HF+BCS wave functions $| q\rangle$ generated by
self-consistent mean-field calculations with a constraint on the axial mass
quadrupole moment
\mbox{$q \equiv \langle q| 2 z^2 - x^2 - y^2 | q \rangle$}.
Dynamic correlations associated with symmetry restorations and fluctuations
in the shape degree of freedom are introduced by particle-number and
angular-momentum projection in the framework of the GCM. Limiting ourselves to
axially symmetric configurations, the final wave function for the correlated
state $| J M;k\rangle$ is given by the superposition of symmetry restored
mean-field wave functions
\begin{equation}
\label{projgcm:wf}
| J M;k\rangle
= \sum_{q} f^{J}_{k}(q)  | J M q\rangle \, ,
\end{equation}
where $k=1$, 2, \ldots labels different collective states for a
given angular momentum $J$. The variable $q$ is the generic notation for the
deformation parameters. The symmetry restored mean-field wave function
is constructed as
\begin{equation}
\label{eq_GCM:08}
 | J M q\rangle
= \frac{1}{{\cal N}_{J,M,q}}\sum_K g^J_K \hat{P}^J_{MK} \hat{P}^Z \hat{P}^N | q \rangle,
\end{equation}
where ${\cal N}_{J,M,q}$ is a normalization factor.
$\hat{P}^{J}_{MK}$ projects out eigenstates of $\hat{J}^2$ and $\hat{J}_z$  with eigenvalues
$\hbar^2 \, J (J+1)$ and $\hbar M$ in the laboratory frame or $\hbar K$ in the intrinsic frame, respectively, whereas $\hat{P}^N$ and $\hat{P}^Z$ project out eigenstates of the particle-number operator for neutrons and protons with eigenvalues $N_\tau$, $\tau = n$, $p$.

The weight functions $f_k^{J}(q)$ and the energies $E_k^{J}$ of the
states $| J M;k\rangle$ are the solutions of the Hill-Wheeler-Griffin (HWG)
equation~\cite{Hil53}
\begin{equation}
\label{eq_GCM:20}
\sum_{q'} \left[ \mathcal{H}^{J}(q,q') - E_k^{J} \, \mathcal{N}^{J}(q,q')
          \right] f_k^{J}(q') = 0 \, .
\end{equation}
The ingredients of Eq.~(\ref{eq_GCM:20}) are  the norm kernel
$\mathcal{N}^{J}(q,q') = \langle J M q | J M q'\rangle$ and the energy
kernel $\mathcal{H}^{J}(q,q')$, which in our calculation is given by a
multi-reference energy density functional that depends on the mixed
density matrix~\cite{Lacroix09}. The formulae used to evaluate the energy
and the norm kernels have been presented in Ref.~\cite{Bender08LH}.

The weight functions $f^{J}_\mu(q)$ in Eq.~(\ref{projgcm:wf}) are not
orthogonal. A set of orthonormal collective wave functions $g_k^{J}(q)$
can be constructed as~\cite{Ring80,Bender03}
\begin{equation}
\label{eq_GCM:30}
g_{Jk}(q)
= \sum_{q'} \big( \mathcal{N}^{J}\big)^{1/2}(q,q') \, f_k^{J}(q')
\, .
\end{equation}
It has to be stressed, however, that the $|g_{Jk}(q)|^2$ quantity does not represent
the probability to find the deformation $q$ in the GCM state
$| J M; k\rangle$. In addition, in the absence of a metric in the
definition of the correlated state $| J M; k\rangle$,
Eq.~(\ref{projgcm:wf}), the values of $g_{Jk}(q)$
for a converged GCM solution still depend on the discretization chosen for
the collective variable $q$, which is not the case for observables such as the
energies or transition probabilities.

Since the correlated states $| JM; k \rangle$ have good angular
momentum, their spectroscopic quadrupole moments $Q_s(J_k)$
\begin{eqnarray}
\label{specQ}
 Q_s(J_k)
 &=& \sqrt{\displaystyle\frac{16\pi}{5}}
 \begin{pmatrix}
 J & 2 & J \\
 J & 0 & -J
 \end{pmatrix}
 \nonumber\\
& & \times \sum_{q^\prime,q}  f^{J_f\ast}_{k_f}(q^\prime) \,
    \langle Jq^\prime || \hat Q_{2}||  Jq\rangle \,
    f^{J_i}_{k_i}(q)
\end{eqnarray}
and the reduced electric quadrupole ($E2$) transition strengths
between them
\begin{eqnarray}
\lefteqn{
B(E2;J_{k_i}\rightarrow J_{k_f})
} \nonumber\\
& = & \dfrac{1}{2J_i+1} \, \Big|
      \sum_{q^\prime,q} f^{J_f\ast}_{k_f}(q^\prime) \,
      \langle J_f q^\prime || \hat Q_{2} || J_i q\rangle \,
      f^{J_i}_{k_i}(q) \Big|^2
\end{eqnarray}
are calculated directly in the laboratory frame without approximation.
The reduced matrix elements entering both expressions are determined as
\begin{eqnarray}
\lefteqn{
\langle J_f q^\prime|| \hat Q_{2}|| J_i q\rangle
}
\nonumber\\
& = &
  \dfrac{(2J_f+1)(2J_i+1)}{2{\cal N}_{J_f,q^\prime}{\cal N}_{J_i,q}}
 \sum_{M=-2}^{+2}
 \begin{pmatrix}
 J_f & 2 &  J_i \\
  0  & M &  -M
 \end{pmatrix} \nonumber\\
 &&\times \int^\pi_0 \! d\theta \, \sin(\theta) \, d^{J_i\ast}_{-M0}(\theta)
 \langle q^\prime| e^{-i\theta\hat J_y} \hat Q_{2M}  \hat P^N\hat P^Z  | q \rangle \, , \nonumber\\
\end{eqnarray}
where $\hat Q_{2M} \equiv e \, r^2 \, Y_{2M}$ is the electric quadrupole
moment operator. To relate the moments in the laboratory frame to
intrinsic deformation parameters, one can  define two dimensionless
quadrupole deformations in the same way as in the static rotor model.
The first one, $\beta^{(s)}$, is related to the spectroscopic quadrupole
moment (for states in a \mbox{$K=0$} band) by:
\beq
\label{beta:s}
\beta^{(s)}(J_k)
= \sqrt{\displaystyle\frac{5}{16\pi}}
  \displaystyle\frac{4\pi}{3ZR^2}
  \left(-\displaystyle\frac{2J+3}{J}\right) Q_s(J_k)
\eeq
and the second, $\beta^{(t)}$, is related to the reduced $E2$ transition strength
\beq
\label{beta:t}
\beta^{(t)}(J_i,k_i\rightarrow J_f,k_f)
= \displaystyle\frac{4\pi}{3ZR^2}
  \sqrt{\frac{B(E2;J_i,k_i\rightarrow J_f,k_f)}
       {e^2\langle J_i 0 20 | J_f0\rangle^2}} \, .
\eeq
The radius $R$ appearing in both expressions is given by $R = 1.2 A^{1/3}$~fm,
with $A$ being the mass number and $\langle J_i 0 20 | J_f0\rangle$
a Clebsch-Gordan coefficient. The nuclear matrix element entering the
electric monopole decay from $| J M; k\rangle$  to $| J M; k^\prime \rangle$
through the emission of conversion electrons is determined by
\begin{equation}
\label{rho:E0}
\rho_{E0}^2(J_k \to J_{k^\prime})
= \left| \frac{\langle J M; k^\prime | e \sum_p r^2_p | J M; k \rangle}{eR^2}
  \right|^2 \, ,
\end{equation}
where $p$ is an index running over all proton single-particle states.
Since the electric transition matrix elements are calculated in the
full model space of occupied single-particle states, there is no need to
introduce effective charges, and we use the  bare charge for protons instead.

The parametrization SLy6~\cite{Cha98} of the Skyrme interaction
and a density-dependent zero-range pairing force with the same
strength of $-1250$~MeV fm$^3$ for
neutrons and protons and a soft cutoff at 5~MeV above and below
the Fermi energy as defined in Ref.~\cite{Rig99}, are adopted in the
construction of mean-field wave functions and the configuration
mixing calculations. As required by the SLy6 parametrization, the
full two-body center-of-mass correction is included in the
variational equations to generate the mean field and in the
calculation of the projected GCM energies.

\section{Results and discussion}
\label{Sec.III}

\begin{figure}[t]
\includegraphics[width=7cm]{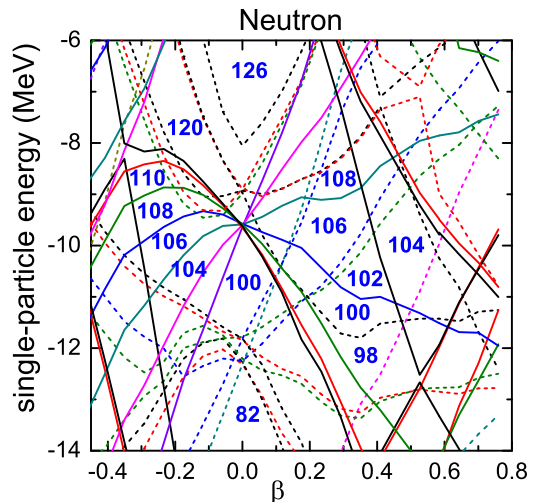}
\caption{(Color online)
Nilsson diagram of the eigenvalues of the single-particle Hamiltonian for
neutrons in $^{190}$Po  as a
function of the axial deformation parameter $\beta$, obtained with the
Skyrme interaction SLy6, see Eq.~(\ref{eq:beta}).
}
\label{spe-n}
\end{figure}

\subsection{General comments}
\label{Sec.III:general}

In mean-field models, self-consistent and non-self-consistent ones
alike, there is an intimate relation between the total binding energy
and the density of levels around the Fermi energy in the Nilsson diagram
of single-particle energies. This is
obvious for non-self-consistent macroscopic-microscopic models, where
the extra binding from shell structure is explicitly calculated as a
``shell correction'' to the total binding energy, and, therefore, can be
easily isolated, but the same mechanism is hidden in the total
energy of self-consistent mean-field models. A lower-than-average density
of single-particle levels around the Fermi energy results in extra binding,
whereas a larger-than-average value reduces binding. Therefore, large
gaps near the Fermi energy in the Nilsson diagram often correspond to
minima in the deformation energy curve, whereas a large bunching of levels
close to the Fermi energy usually corresponds to barriers between such
minima. The Fermi energy does not have to be located exactly at the gap,
but might be slightly below or above it. With increasing distance of the
Fermi energy from the gap, however, the shell effect can be expected to
become less pronounced. In any event, shape coexistence is the fingerprint
of a variation of shell structure around the Fermi energy that opens up
or closes gaps with deformation.

To have near-degenerate collective states with the same quantum numbers,
but different deformations, there has to be a mechanism that prevents
their mixing. As states of different deformations are \textit{a priori}
not orthogonal, the diagonalization of the Hill-Wheeler-Griffin equation
(\ref{eq_GCM:20}) can easily produce
collective wave functions spread over a large range of deformations.
Large non-diagonal energy kernels will amplify this effect.
Therefore, the sole presence of a large barrier
between the minima is not sufficient to prevent mixing. To achieve a weak
coupling between mean-field configurations, the matrix elements of the
overlap and the energy kernels between them must be small.
A mechanism leading to such suppression of the matrix elements
is provided by  intruder states; i.e.\, single-particle levels that are
downsloping with deformation from the next major shell and,
therefore, have a parity opposite to that of the levels at
the Fermi energy. Assuming conserved parity and neglecting pairing
correlations, two HF states with a different number of
positive-parity and negative parity single-particle levels do not overlap
and are weakly coupled in the GCM. This is no longer the case when
pairing correlations are taken into account, but one can expect that the
overlap between the mean-field wave functions remains small.

The detailed results of this extensive study are available
in the supplemental files {\tt tab1.dat} and {\tt tab2.dat}.
They include excitation energies, charge radii, spectroscopic quadrupole
moments, electric monopole and quadrupole transition strengths for the
low-lying states that were computed for total angular momentum values
up to $J=10\hbar$, in the neutron-deficient $^{176-194}$Hg, $^{180-194}$Pb,
$^{186-210}$Po and $^{194-204}$Rn isotopes. We focus here on the
evolution of low-lying states with neutron and proton number and discuss
the similarities and differences between these four isotopic chains.

\subsection{Nilsson diagrams}

\begin{figure}[t]
\includegraphics[width=7cm]{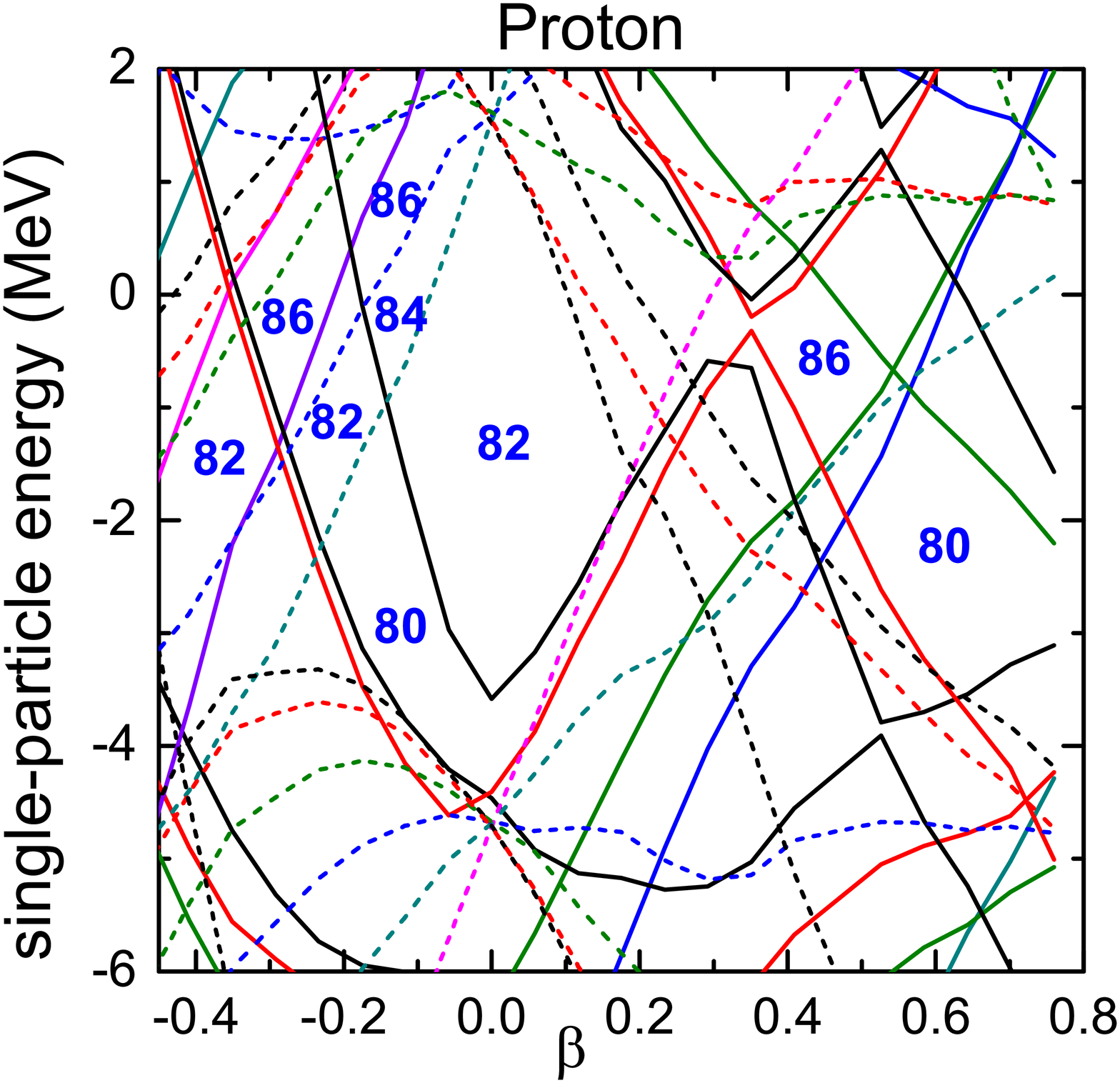}
\caption{(Color online) Same as Fig.~\ref{spe-n}, but for the protons.
}
\label{spe-p}
\end{figure}

Figures~\ref{spe-n} and ~\ref{spe-p}
display the Nilsson diagram of neutron and proton
single-particle energies as a function of a dimensionless quadrupole
deformation parameter $\beta$. This parameter is related to the intrinsic mass
quadrupole moment $q$ of the HF+BCS states by the relation
\begin{equation}
\label{eq:beta}
\beta
= \sqrt{\frac{5}{16 \pi}} \frac{4 \pi}{3 A R^2} q \, .
\end{equation}
There is a large number of gaps visible in the Nilsson diagram for
neutrons. First, there is a spherical one at $N=100$, below the
$1i_{13/2^+}$ intruder level. There are many deformed gaps of slightly
varying size on the oblate side for every even neutron number from
$N=102$ to $N=110$ at increasingly large deformations up to
$\beta \simeq -0.2$. These gaps are all located between the spreading magnetic
substates of the $1i_{13/2^+}$ shell. There are also gaps of varying size
for every even neutron number between $N=98$ and $N=108$ on the
prolate side.  Due to the downsloping $K = 1/2$ level from the $2g_{9/2^+}$
shell above and the upsloping levels from the $1h_{9/2^-}$ shell  below,
however, they correspond to configurations with a different number of
occupied intruder levels. Because of several level crossings,
the deformations of the gaps on the prolate side are not ordered
according to $N$. The largest ones are  $N=106$ for $\beta \simeq 0.3$ and
$N=104$ for $\beta \simeq 0.5$, respectively.

For the protons, there is a large spherical shell gap at $Z=82$, a gap
at $Z=80$  extending from sphericity to an oblate shape with $\beta$ values
down to $-0.2$ as well as a smaller oblate $Z=82$ gap at $\beta \simeq -0.25$.
At normal deformation, there are no significant proton gaps on the
prolate side that come close to the Fermi energy, except for a small
$Z=86$ one at \mbox{$\beta \simeq 0.5$}.

At the deformation corresponding to the deformed neutron gaps, the
downsloping high-$j$ proton intruder orbitals have dropped to energies
close to the Fermi energy, in most cases even well below, and, thus, have
a sizable occupation. For this reason, some single-particle orbitals
composing the corresponding deformed mean-field wave function are very
different from those of the spherical configuration and the overlap
between both is small. As discussed above, a small overlap creates
favorable conditions for a decoupling between configurations corresponding
to different shapes.

We have checked that the Nilsson diagrams of both protons and
neutrons change only marginally for the nuclei covered in the present
study.

\subsection{Systematics of deformation energy curves}

\begin{figure}[t!]
\includegraphics[width=7cm]{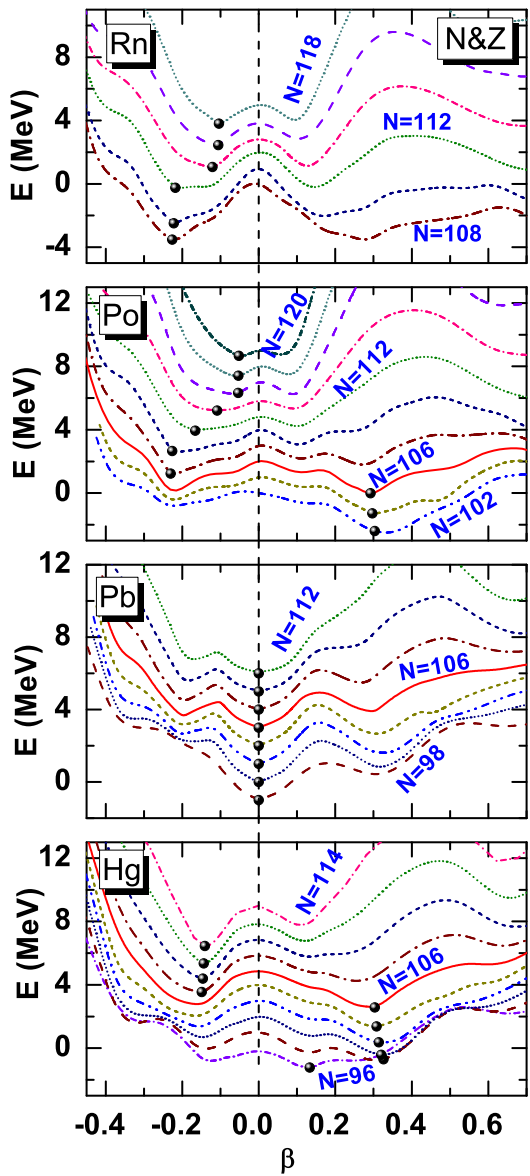}
\caption{
\label{PEC:NZ}
(Color online) Energy curves for the particle-number-projected
HF+BCS states ($N$\&$Z$) for $^{176-194}$Hg, $^{180-194}$Pb, $^{186-204}$Po,
and $^{194-204}$Rn as a function of the intrinsic quadrupole deformation
parameter $\beta$. All energies are normalized to the spherical shape
($\beta=0$), but with an additional energy shift of 1 MeV between two
neighboring isotopes. The lowest configuration of a given nucleus in
the deformation region of interest is indicated with a bullet.
}
\end{figure}

The energy curves obtained after projection onto particle numbers
are presented in Fig.~\ref{PEC:NZ}. The mean-field ground states for all
neutron-deficient Pb isotopes are spherical. The same result has been
obtained in calculations using the parametrization D1S of the Gogny
force~\cite{Guzman04}.
However, it can be modified by a lower pairing strength~\cite{Ben06a}
or the use of a different pairing functional~\cite{Car12a}. For
all Pb isotopes, except $^{180-182}$Pb, the curves also display an oblate
minimum with $\beta \simeq -0.20$. The excitation energy of this minimum
decreases from 0.76~MeV to 0.50~MeV
 for $N$ going from 112 to 108, and increases again up to 1.63~MeV when
decreasing the neutron number further down to 102. The well depth follows
a similar evolution. It starts from 0.46 MeV for $N=112$, is maximal at
0.93~MeV for $N=108$ and decreases again down to 0.49~MeV for $N=102$.
On the prolate side, the energy curves present an inflexion point at
\mbox{$\beta \simeq 0.30$} for $N\ge110$ and a minimum for lower $N$
values down to 100. The excitation energy of this minimum decreases
from 1.47~MeV in $^{190}$Pb down to 0.61~MeV in $^{184}$Pb and
raises up again to 0.82~MeV in $^{182}$Pb.

The deformation of the prolate minima of the Pb isotopes follows
closely the neutron gaps in the Nilsson diagram. Their deformation
is \mbox{$\beta_2 \simeq 0.32$} for $N=100$ increases slightly at $N=102$,
and then decreases up to $N=106$. For $108 \leq N \leq 112$, this prolate minimum
disappears as the Fermi energy approaches a region with densely bunched
levels in the Nilsson diagram.

There are no spherical minima for the Hg, Po and Rn isotopes discussed here.
The  minima of the deformation energy curves for these nuclei are not always
in a one-to-one correspondence with the gaps in the Nilsson diagram. This
is because the gap for one nucleon species is often located
at a deformation corresponding to densely bunched levels for the other.
Hence, the minima correspond to the most favorable compromise between the shell
effects for neutrons and protons.

The ground states that we obtain for the Hg isotopes are oblate above
$N=106$ and prolate below this neutron number. At $N=106$, the oblate and prolate minima
are nearly degenerate. The energy curves are very shallow for $N=98$ and
$96$. The deformation of the oblate minimum corresponds to the $Z=80$ proton
gap for all values of $N$, whereas the prolate minima of the isotopes
with \mbox{$N \geq 98$} correspond to the many gaps obtained
in the neutron Nilsson diagram around \mbox{$\beta \simeq 0.3$}. For
the soft \mbox{$N = 96$} isotope, the prolate minimum is shifted to a
weakly deformed configuration with \mbox{$\beta \simeq 0.13$}, close in
energy with the weakly deformed oblate minimum.
The oblate minimum is very rigid for $N=114$ and
becomes softer with decreasing $N$. Such a behavior can be related to the
neutron single-particle levels close to the Fermi energy. For $N \geq 108$,
the density of neutron levels is high and their energy is rapidly increasing
when $\beta$ becomes more negative. The situation changes at $N=106$
and below, where the density of neutron levels is low and the energy
varies more slowly with deformation.

\begin{figure}[t!]
\includegraphics[width=9cm]{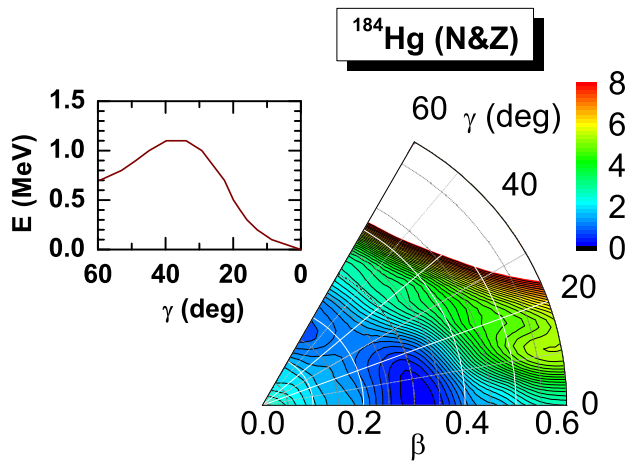}%
\vspace{-0.6cm}%
\caption{\label{Hg184_triaxial}
(Color online) Quadrupole deformation energy surface for
$^{184}$Hg, normalized to the absolute minimum and projected on
particle numbers. Each contour line is separated by 0.2~MeV. The
inset shows the energy as a function of $\gamma$ deformation
along the path joining the two axial minima.
}
\end{figure}

The energy curves of the Po isotopes evolve in a way similar to those
of the Hg isotopes. In particular, the transition between oblate and
prolate ground states appears at the same neutron number. However,
there are differences in the details. The oblate minimum is always shallower,
and the oblate deformation of the ground state increases steadily with
decreasing $N$. For $N=108$, which corresponds to the lightest Po isotope
with an oblate ground state, the deformation parameter $\beta$ is
significantly larger for Po than for Hg. By contrast, for
\mbox{$102 \leq N \leq 106$} the prolate ground states have a smaller
deformation for Po than for the corresponding Hg isotopes.

The energy curves of the Rn isotopes evolve in a way similar to those
of the Po isotopes, but with deeper oblate and prolate minima.

Our calculations are limited to axial deformations.
One cannot exclude, however, that in some cases the minima appearing
in the axial deformation energy curves actually correspond to saddle points
of the energy surface in the full $\beta$--$\gamma$ plane. Several
scenarios are possible. First, the axial minima can be separated by a barrier
with the energy rising all the way from both sides, such that they are
true minima. In Ref.~\cite{Bender04}, we have checked that this is the
case for the spherical, prolate and oblate minima found in the
\nuc{182-194}{Pb} isotopes studied there. But there could be cases where
there is no barrier between the axial minima. A second possibility is that
the energy rises monotonously from \mbox{$\gamma = 0^\circ$} to~$60^\circ$.
Then, the higher-lying minimum is in fact a saddle point. This happens for many
well-deformed nuclei, and also for many $\gamma$-soft ones. A third case
occurs when there is a triaxial minimum between the two axial ones.
In this case, depending on the appearance of barriers, one or even both
minima in the axial energy curve might be saddle points.

\begin{figure}[t!]
\includegraphics[width=9cm]{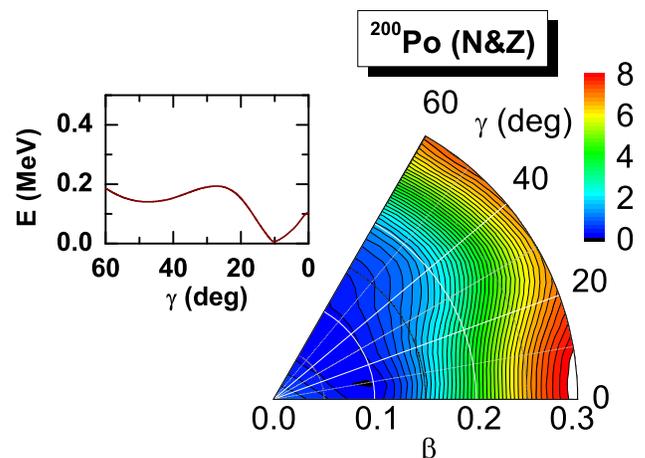}%
\vspace{-0.6cm}%
\caption{\label{Po200_triaxial}
(Color online) Same as Fig.~\ref{Hg184_triaxial} for $^{200}$Po.}
\end{figure}

The Gogny interaction D1s very often gives deformation energy surfaces
very similar to those obtained with the Skyrme interaction SLy6
used here. A systematic survey of energy surfaces in the $\beta$--$\gamma$
plane using D1s can be found in~\cite{bruyeres}. For the
nuclei studied here, the study of ~\cite{bruyeres} indicates that  minima found in a calculation
restricted to axial shapes have to be suspected to be saddle points when
the ground state and the secondary minimum in the axial energy curve
are nearly degenerate and are obtained for similar values for $|\beta|$.

Two examples of particle-number projected energy surfaces in the
full $\beta$--$\gamma$ plane obtained with SLy6 are displayed in
Fig.~\ref{Hg184_triaxial} for $^{184}$Hg and Fig.~\ref{Po200_triaxial}
for $^{200}$Po, respectively. The variation of the energy as a function
of $\gamma$ for a path going from the oblate to the prolate minimum is
shown as an inset in the figures.

Fig.~\ref{Hg184_triaxial} illustrates the case where there is small
barrier between the two minima. For $^{200}$Po, the absolute minimum
is obtained for a non-axial configuration, with \mbox{$\beta \approx 0.1$}
and \mbox{$\gamma \approx 10^{\circ}$}.
This nucleus is a rare case where both minima found in an
axial energy curve actually correspond to saddles in the full
$\beta$--$\gamma$ plane \cite{Mol06a}. The prolate saddle is at almost the
same $|\beta|$ value as the nearby triaxial minimum and is located just about
100~keV higher, whereas the oblate saddle
is at somewhat smaller $|\beta|$ values and higher excitation energy.
The heaviest Po and Rn studied here can be expected to have deformation
energy surfaces of similar $\gamma$-soft topography, although not
necessarily with a triaxial minimum.

\begin{figure}[t!]
\includegraphics[width=7cm]{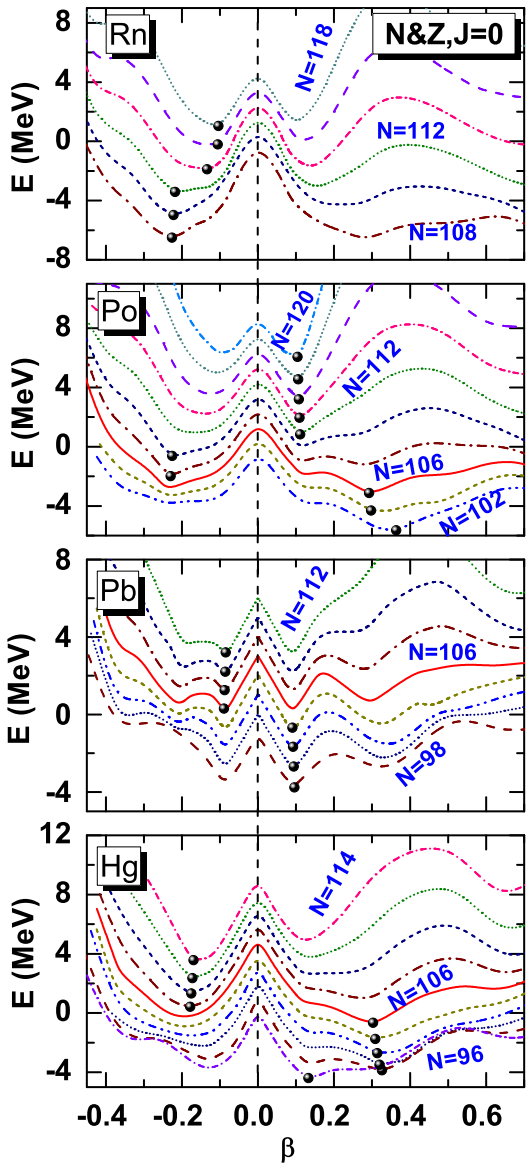}
\caption{\label{PEC:J0}
(Color online)
Same as Fig.~\ref{PEC:NZ}, but with projections onto
both particle numbers and angular momentum $J=0$.
}
\end{figure}

\subsection{Effects of angular momentum projection on energy curves}

\begin{figure}[t!]
\includegraphics[width=6cm]{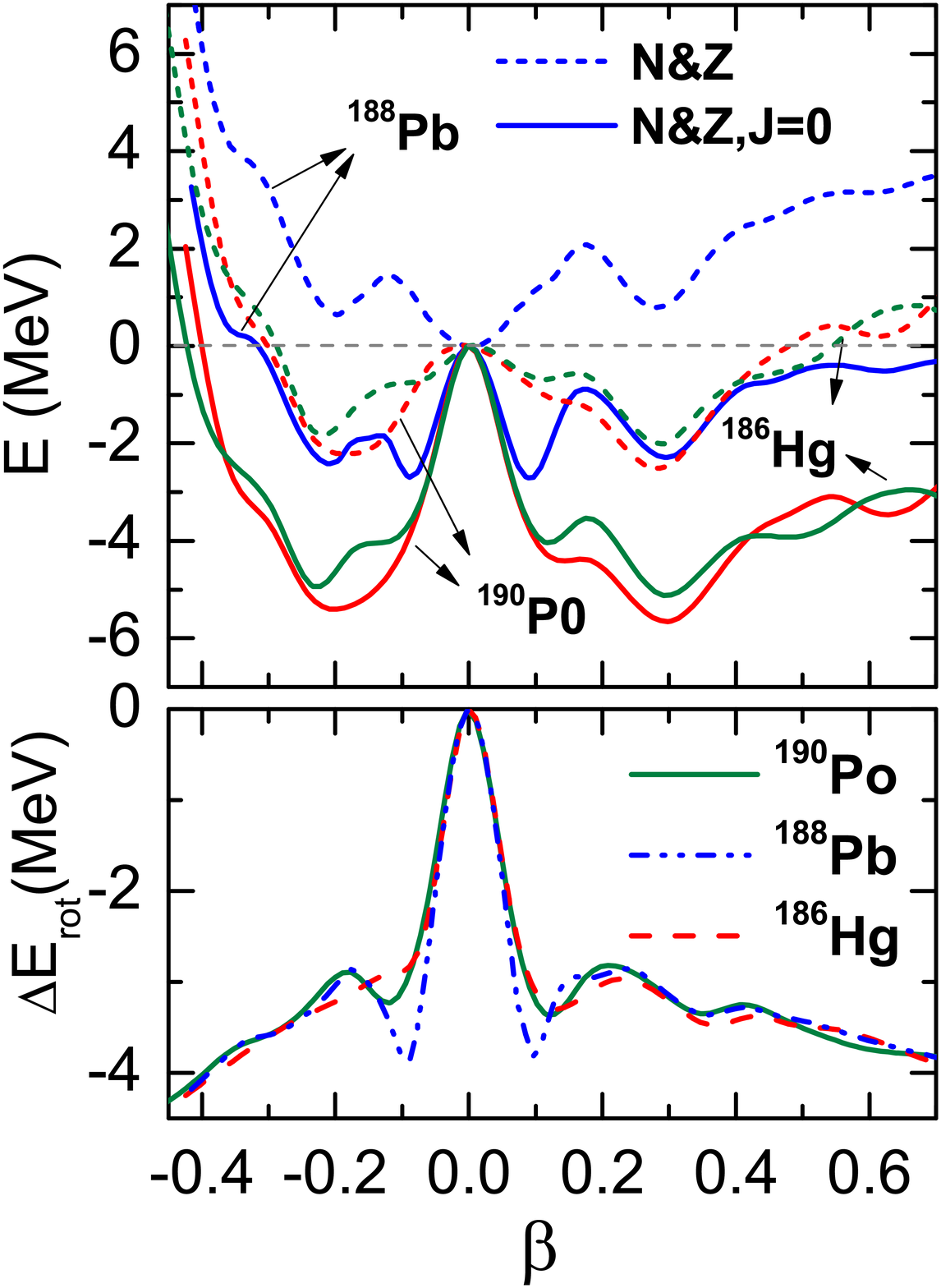}
\caption{\label{energy:correction}
(Color online)
Effect of projection on $J=0$ on the topography of
deformation energy curves at small deformation. Top panel:
energy curves after projection on particle numbers
($N\&Z$) and on angular momentum $J=0$ as a function of $\beta$
for$^{186}$Hg, $^{188}$Pb, and $^{190}$Po.
Lower panel: The corresponding energy gained from projection on $J=0$;
$\Delta E_{\text{rot}}(\beta)= E_{J=0}(\beta)-E(\beta)$.
}
\end{figure}

The energy curves projected on $J=0$ are given in Fig.~\ref{PEC:J0}.
The energies are drawn at the intrinsic quadrupole moment (or the
$\beta$-value) of the mean-field wave function that is projected.
This is the most convenient way to plot the results obtained after
projection on  angular momentum. However, it has to be kept in mind that,
after angular-momentum projection, it is only at large deformation and
for \mbox{$J > 0$} that the $\beta$ value of Eq.~(\ref{eq:beta}) provides a rough
estimate for the values obtained through Eqs.~(\ref{beta:s}) or~(\ref{beta:t})
from observable quadrupole moments. At small deformations, the intrinsic
deformation does not have a relation to an observable. Therefore,
the interpretation of these energy curves requires some
caution~\cite{Bender04,Guzman04}.
The energy \mbox{$\Delta E_{\rm rot} \equiv E_{J=0}(\beta)-E(\beta)$} gained
from the restoration of rotational symmetry  is displayed on the bottom
panel of Fig.~\ref{energy:correction} for $^{186}$Hg, $^{188}$Pb, and
$^{190}$Po.  The spherical configuration is purely a \mbox{$J=0$}
configuration; hence, projection on \mbox{$J=0$} does not bring any gain
of energy. A slight deformation of the mean-field is sufficient to introduce higher $J$
components (or, in a shell-model language, particle-hole excitations) in the
mean-field wave function. Projection on \mbox{$J=0$} gives then an energy
gain, which at small deformation is almost symmetrical around
the spherical point and increases rapidly (in absolute value) to reach
about 3~MeV around $|\beta | \approx 0.1$. At larger deformation, the
energy gain still increases further, but at a slower rate, as illustrated
in Fig.~\ref{energy:correction}.
For nuclei with deformed minima, this generic behavior usually makes the
energy curve more rigid at small deformation values and softens it at larger
ones. For nuclei with a spherical minimum in the mean-field energy
surface, such as the Pb isotopes, the minimum  is shifted to small absolute
values of the deformation $\beta$ for all values of $\gamma$,
forming a kind of ``Mexican hat" around the spherical point.

Neither the mean-field wave functions nor the wave functions
obtained for a given set of quantum numbers after symmetry restoration
are orthogonal. The overlap between \mbox{$J=0$} wave functions is large
for all small deformation values of the mean-field. It turns out that the
states ``at the bottom of the mexican hat" are not only
almost degenerate, but also have very large overlaps close to one,
meaning that after projection they describe the same physical state.
We will call a state in this very particular minimum a "correlated
spherical state". This state has the property to appear
only for \mbox{$J=0$} values, but not for higher angular momenta, and therefore,
it is not the head of a rotational band. This allows to distinguish the
correlated spherical states from "truly deformed" minima at small
deformation that appear at all $J$ and lead to the appearance of a
rotational band.
In any event, at small deformation, the labeling of projected wave
functions by a mean-field deformation has a very limited meaning.

The main findings for the effect of projection on $J=0$ on the
energy curves of the nuclei discussed here are:
\begin{itemize}
\item
Using the definition of a correlated spherical state, the ground state
is spherical for all Pb isotopes. A prolate minimum around
\mbox{$\beta \simeq 0.3$} is obtained for the lightest isotopes up to
$N=106$. The oblate minimum that was obtained before projection on
angular momentum for isotopes above $N=106$ is not clearly visible
anymore, as it merges with the spherical minimum.

\item
For the Hg, Po and Rn isotopes, angular momentum projection does not
modify the topography of the energy curves significantly. Both the
prolate and the oblate minima become more bound with respect to the
energy of the spherical point. For some nuclei, the minimum is also
shifted to slightly larger deformation.

\item
For Po isotopes with \mbox{$112 \leq N \leq 120$}, the absolute minimum
of the  energy curves moves from an oblate to a prolate shape after projection on $J=0$. These Po nuclei are the only ones where projection results in a change of the sign of the deformation.
\end{itemize}
When triaxial shapes are included, angular momentum projection shifts
the deformed minima from axial shapes to slightly non-axial
ones~\cite{Bender2008,Yao10,Rodriguez10}, in the same manner as the
spherical minimum is shifted to slightly deformed intrinsic shapes.
For light nuclei, symmetry-restored GCM calculations including triaxial
shapes have been performed and this shift does not qualitatively change
the interpretation of the minima. Because of their high computational
cost, such calculations have not been performed yet for nuclei as heavy
as the ones discussed here within an energy density functional framework.

While the mean-field ground states are at normal deformation for all
nuclei considered here, we have to note that after angular momentum
restoration, the ground states of $^{186,188}$Po and $^{194,196}$Rn
correspond to the projection of a superdeformed configuration beyond
the range of deformations displayed in Fig.~\ref{PEC:J0}. A similar
result for the lightest Po and Rn isotopes is also obtained when using the
Gogny D1s force and adding a rotational correction \cite{bruyeres}.
This finding is the consequence of the energy gain from angular momentum
projection growing with deformation, cf.\ Fig.~\ref{energy:correction}.
It is in contradiction with the data
and is an artefact of the low
surface energy coefficient $a_{\rm surf} = 17.7$~MeV of the
parametrization SLy6. We have checked that with the SLy4 parametrization, which has a slightly
larger surface energy coefficient of $a_{\rm surf} = 18.4$~MeV, the
deformation energy curves are stiffer such that the superdeformed
minima remain above the normal deformed ones after projection. On the
other hand, at the mean-field level, the relative excitation energies
of the various minima in the Pb region are much better described by
SLy6, as, overall, SLy4 gives excitation energies that are too large. A similar
result has been found for $^{240}$Pu in Ref.~\cite{Bender2004}.
This points to the need to fit dedicated parametrizations for
beyond-mean-field calculations in the future.
We have limited our configuration mixing calculations to deformations
up to \mbox{$\beta \approx 0.6$}. We have checked that the normal-
and superdeformed states are sufficiently decoupled that the low-lying states
at normal deformation discussed here are not affected by
using these restricted configurations.

\subsection{Systematics of low-lying $0^+$ states}
\subsubsection{Collective wave functions}

The collective wave functions $g_k^{J}(q)$, as defined by
Eq.~(\ref{eq_GCM:30}), are spread over a large range of deformed
mean-field wave functions. However, it is often still possible
to classify them as spherical, oblate or prolate by looking at the dominant
configurations. Such a classification is greatly helped by studying
the mean deformation of the mean-field components of a collective
wave function defined by
\begin{equation}
\label{beta:average}
\bar\beta_{Jk}
\equiv \sum_q \beta(q) \, |g_{Jk}(q)|^2 \, .
\end{equation}
This quantity has to be taken with a grain of salt. First, because of
the non-orthogonality of the basis of projected mean-field states,
$|g_{Jk}(q)|^2$ does not represent the probability to find a given
mean-field state $| q\rangle$ in the projected GCM state $| JM;k\rangle$.
Second,  $\bar\beta_{Jk}$ can not be expected in general to correspond
to a deformation deduced from electromagnetic transitions or spectroscopic
moments. Still, this quantity turns out to be useful when analyzing
the mixing of oblate, spherical and prolate wave functions.
For simplicity, we will label the collective states as oblate,
prolate or spherical when the mean deformation $\bar\beta_{Jk}$ has a
value close to the deformation of the mean-field state with the largest
weight in  the collective wave function. When this is not the case, the
state will be denoted as "spread".

\begin{figure}[t!]
\includegraphics[width=8.2cm]{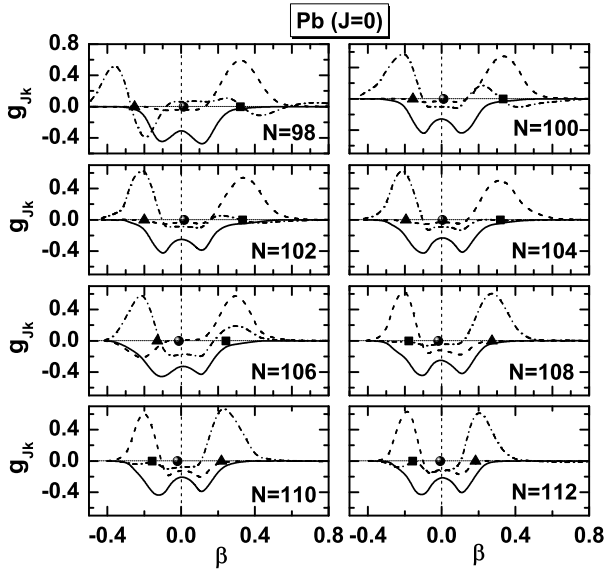}%
%\vspace{-0.4cm}%
\caption{\label{wfs1:J0}
Collective wave functions for the first two and some
selected third $0^+$ states in $^{180-194}$Pb. The mean deformations
$\bar\beta_{Jk} \equiv \sum_q \beta(q) \, |g_{0k}(q)|^2$  are indicated
with the symbols $\bullet (0^+_1)$, $\blacksquare (0^+_2)$, and
$\blacktriangle (0^+_3)$.
}
\end{figure}

\begin{figure}[t!]
\includegraphics[width=8.2cm]{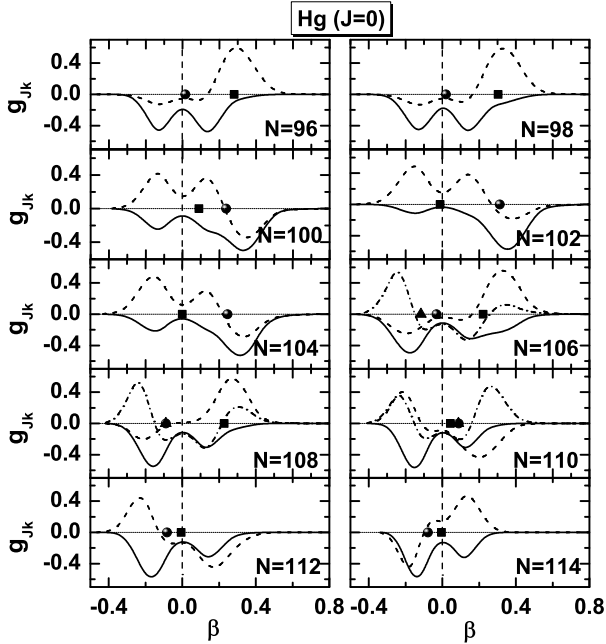}%
%\vspace{-0.6cm}%
\caption{\label{wfs2:J0}
Same as Fig.~\ref{wfs1:J0}, but for $^{176-194}$Hg.
}
\end{figure}

\begin{figure}[t!]
\includegraphics[width=8.2cm]{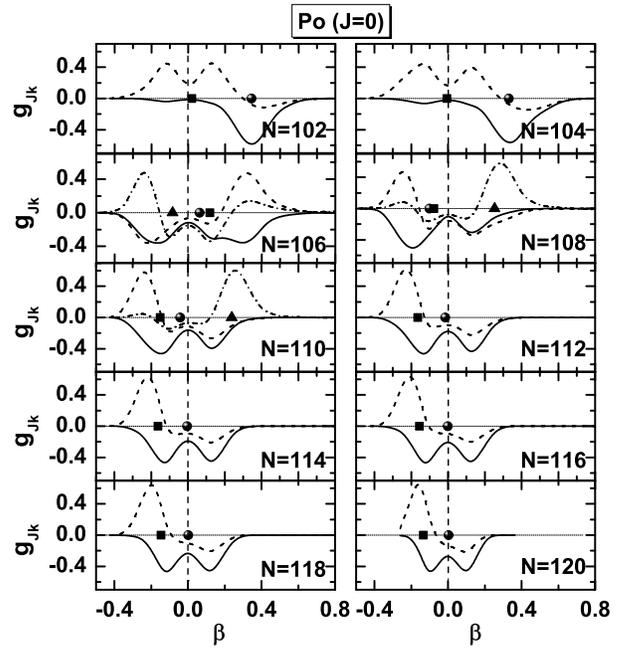}%
%\vspace{-0.6cm}%
\caption{\label{wfs3:J0}
Same as Fig.~\ref{wfs1:J0}, but for $^{186-204}$Po.
}
\end{figure}

\begin{figure}[t!]
\includegraphics[width=8.2cm]{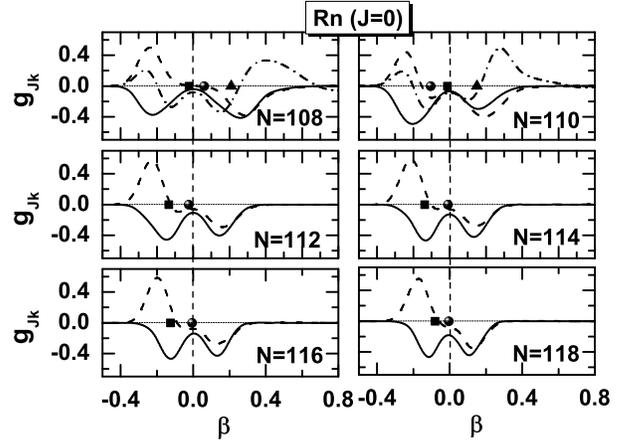}%
%\vspace{-0.6cm}%
\caption{\label{wfs4:J0}
Same as Fig.~\ref{wfs1:J0}, but for $^{194-204}$Rn.
}
\end{figure}

The collective wave functions $g_{Jk}(q)$, Eq.~(\ref{eq_GCM:30}),
of the lowest $J=0$ states are plotted in Figs.~\ref{wfs1:J0}-\ref{wfs4:J0} as a function of the deformation
$\beta$ for $^{180-194}$Pb, $^{176-194}$Hg,$^{186-204}$Po
and $^{194-204}$Rn, respectively. The third $0^+$ is displayed only when
it presents a special interest. The $\bar\beta_{0k}$-value is indicated
for each state by the symbols $\bullet (0^+_1)$, $\blacksquare (0^+_2)$,
and $\blacktriangle (0^+_3)$, respectively.

For all Pb isotopes, the GCM ground state is predominantly spherical,
in agreement with the data. The two lowest excited states are dominated
by either oblate or prolate components.  As we shall see below, the
prolate state is lower in energy for the lighter isotopes
up to $N=106$ and the oblate one is lower above this value while they are nearly
degenerate for $N=106$. Although they are not separated by a sizable
barrier, the spherical and oblate states are well separated. The
occupation of intruder single-particle states in the oblate configuration
seems to prevent a mixing. Indeed, at $\beta\simeq-0.15$, the downsloping
$K=7/2$ intruder orbital from the $2f_{7/2^-}$ shell above the spherical
$Z=82$ shell closure crosses the upsloping $3s_{1/2^+}$ orbital. In the
same way, the prolate state does not mix significantly with the spherical
and oblate ones as its dominant contributions have a large occupation of
several downsloping intruder levels on the prolate side.

The mean deformation $\bar\beta_{01}$, Eq.~(\ref{beta:average}),
for the ground states of $^{176}$Hg and $^{178}$Hg displayed in
Fig.~\ref{wfs2:J0} is close to zero. Indeed, the small value of
$\bar\beta_{01}$ is the result of a cancelation between the contributions
of oblate and prolate configurations.
This large mixing of different shapes can be related to
flat energy curves given for these two isotopes in Fig.~\ref{PEC:NZ}.
For both isotopes, the first excited $0^+$ state is prolate.

In fact, from the discussion of triaxial energy surfaces above, one can
assume that both of these nuclei are soft in the $\gamma$ degree
of freedom that is not explicitly considered here. However, we recall that
in symmetry
restored GCM, prolate and oblate configurations are directly mixed by
the Hill-Wheeler-Griffin equation, Eq.~(\ref{eq_GCM:20}), independently
of the height of a barrier on the axial path between them.
Therefore, in an axial GCM calculation, as done here, one cannot distinguish
the wave function of a $\gamma$-soft nucleus from that resulting
from large mixing of two minima separated by a barrier.

For the Hg isotopes \mbox{$100 \leq N \leq 104$}, the prolate configuration is
the ground state in contradiction with the data, as we shall see
below, and the first excited $0^+$ state is a spread
state. The balance between oblate and
prolate configurations in the ground state is reversed above $N=104$,
with a larger weight on the oblate side for the heaviest isotopes
represented in the Figure. The first excited $0^+$ state is predominantly
prolate for $N=96$, $98$, $106$ and $108$ and has a large spreading above 108.
The second excited state is oblate for $N=106$ and $108$.

For the Po isotopes, the ground states for $N=102$ and $104$ are the
only ones to be dominated by prolate configurations. The ground states
of the heavier isotopes have a large spreading.
Figure~\ref{Po200_triaxial} indicates that for these isotopes, the
almost equal mixing of prolate and oblate configurations results from the
$\gamma$ softness of their energy surface. The mean
deformation $\bar\beta_{Jk}$ of excited states for $102\leq N \leq 106$
is small, reflecting the  nearly equal weight of oblate and prolate
configurations. From $N=108$ to the heaviest Po isotopes studied here,
the first excited $0^+$ state is dominated by oblate components, whereas
the second excited one is prolate for $N=108$ and $110$. Experimentally,
the ground states of the Po isotopes are interpreted as being spherical
down to $N=112$. The lighter isotopes are inferred to posses deformed
ground states~\cite{Heyde11}.

The collective wave functions of the Rn isotopes behave in a
way similar to those of their Po isotones.

\begin{figure}[t!]
\includegraphics[width=9cm]{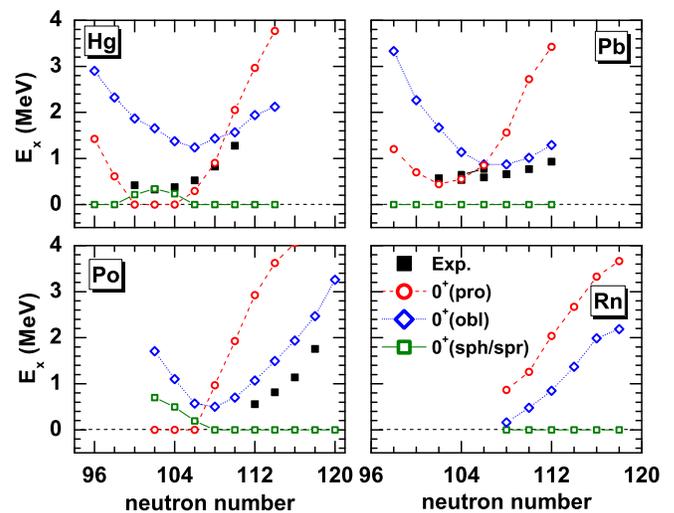}
\caption{\label{EJ0}
(Color online)
Systematics of the excitation energy of the first three $0^+$ states
in the Hg, Pb, Po and Rn isotopes. Lines connect the states with similar
dominant configurations. The available experimental
data~\cite{NNDC,Page11,Elseviers11} for the low-lying $0^+$ states are
plotted for comparison.
}
\end{figure}

\subsubsection{Excitation energies}

The evolution of the energy of the first excited $0^+$ states
is compared to the experimental data in Fig.~\ref{EJ0}. The main
findings are as follows:
\begin{itemize}
\item
Going from $N=120$ down to $N=108$,
the $0^+_2$ and $0^+_3$ excitation energies
decrease rapidly in all the four isotopic chains, in agreement with the data.

\item
As discussed above, the  states  $0^+_2$ and $0^+_3$ in Pb isotopes all
have a well defined shape, the $0^+_2$ level being oblate above $N=106$
and prolate below. The oblate and prolate configurations cross at $N=106$;
in this nucleus the excitation energies of the $0^+_2$ and $0^+_3$ states
are quite close and below 1~MeV.

\item
The evolution of the energy of the  $0^+_2$ state in Hg isotopes is
in agreement with the data. In our calculations, this state can be called
prolate only for $N = 96$, 98, 106 and~108; for all other isotopes, it has
a very spread character. Also, for \mbox{$100 \leq N \leq 104$}, the
agreement between calculations and data is fortuitous, since our calculations
do not reproduce the shape of the ground state deduced from the data.

\item
For Po isotopes, the calculations overestimate the excitation energy of
the $0^+_2$ level but reproduce the slope of the variation of the
energy with $N$. For all isotopes for which data exist, our calculations
predict the $0^+_2$ state to be oblate.

\end{itemize}

\begin{figure}[t!]
\includegraphics[width=9cm]{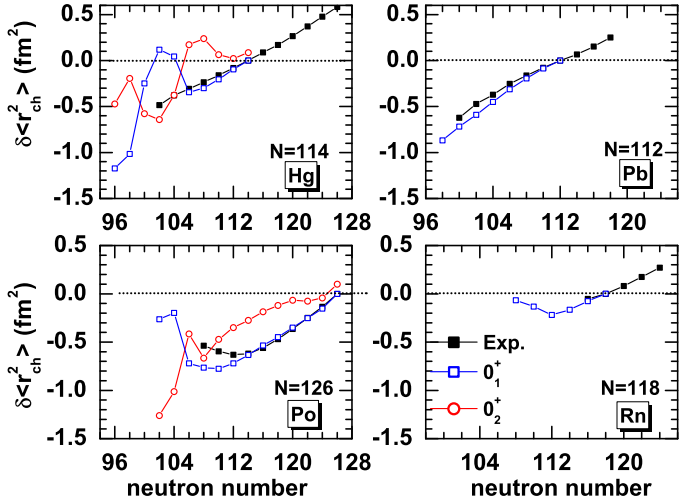}
\caption{\label{isotope-shifts}
(Upper panel) Variation of the charge radii
$\delta\langle r^2_{\rm ch}\rangle$ for the lowest $0^+$ states in Hg
(normalized to the ground state (g.s.) of $^{194}$Hg), Pb (normalized to
the g.s.\ of $^{194}$Pb), Po (normalized to the g.s.\ of $^{210}$Po) and
Rn (normalized to the g.s.\ of $^{204}$Rn) isotopes, compared to the
the experimental data for ground states taken from
Refs.~\cite{Angeli04,Coc11}.
}
\end{figure}

\subsubsection{Variation of mean-square charge radii}

The variation of the calculated mean-square charge radii is compared in
Fig.~\ref{isotope-shifts} with the values deduced from the measured
isotopic shifts. Our results for the Pb isotopes obtained with the SLy6 and
SLy4 Skyrme parametrizations, each used with a different pairing strength,
have already been compared with the data in Ref.~\cite{Witte07}. In
Ref.~\cite{Coc11}, results obtained with the SLy4 parametrization and
two different values of the pairing interaction strength have
been confronted with data. In both cases, it was observed that charge
radii are extremely sensitive to the amount of mixing of mean-field
wave functions corresponding to different shapes, which in turn is
very sensitive to the pairing strength.

A sudden increase of the charge radii is obtained at $N=104$ in our
calculation for the Hg isotopes. It is related to the shape transition
seen in Fig.~\ref{PEC:NZ} and \ref{wfs2:J0} that is, however, not corroborated
by the data. By contrast, a jump of similar size has been observed in
odd-$A$ Hg \cite{Bonn72,Raman01}. This observation indicates that two states based on
either prolate or oblate shapes indeed coexist but that our calculations
do not predict correctly their relative position. Above \mbox{$N=104$},
the ground-state wave functions are weakly oblate, but become strongly
peaked at prolate shapes for \mbox{$100 \leq N \leq 104$}, as shown in
Fig.~\ref{wfs2:J0}. However, this onset of large prolate deformation
in the ground state is in contradiction with the experimental data.
Clearly, the dominant configuration of the ground state should not
change much down to at least $N=102$.

In the Rn isotopes, the kink at $N=112$ in the calculated values
of $\delta\langle r^2_{\rm ch}\rangle$ is due to the onset of weakly
oblate deformations in the ground state at $N=110$, cf.\
Figs.~\ref{wfs2:J0} and \ref{EJ0}.

\begin{figure}[t!]
\includegraphics[width=9cm]{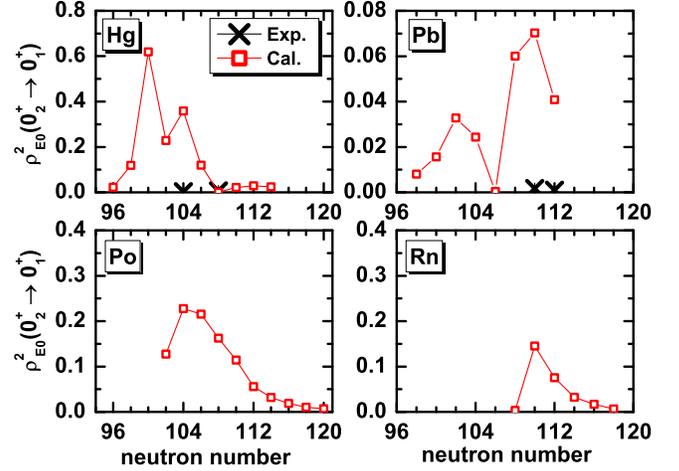}
\caption{ \label{BE0}
(Color online)
Electric monopole transition strengths $\rho^2_{\rm E0}$
for the transitions between the first excited $0^+$ state and the ground
state (c.f.\ Fig.~\ref{EJ0}) in Hg, Pb, Po and Rn isotopes.
Experimental data are taken from Ref.~\cite{Kibedi05}.
}
\end{figure}

\subsubsection{Monopole transition strength}

Figure~\ref{BE0} displays the $E0$ strength $\rho^2_{E0}$,
Eq.~(\ref{rho:E0}), between the $0^+_2$ and the $0^+_1$ states. This electric
monopole transition strength is correlated to the size of the deformation
and to the amount of mixing between configurations corresponding to
different shapes~\cite{Heyde88}. In general, large $E0$ strength arises
from a strong mixing between states with different radii.

Compared to the very sparse experimental data that exist for some Hg,
Pb and Po isotopes, our calculated $\rho^2_{E0}$ values are too large by
an order of magnitude. According to Ref.~\cite{Heyde88}, this
indicates a too stong mixing of configurations with different deformation
in at least one of the two $0^+$ states.

\begin{figure}[t!]
\includegraphics[width=7cm]{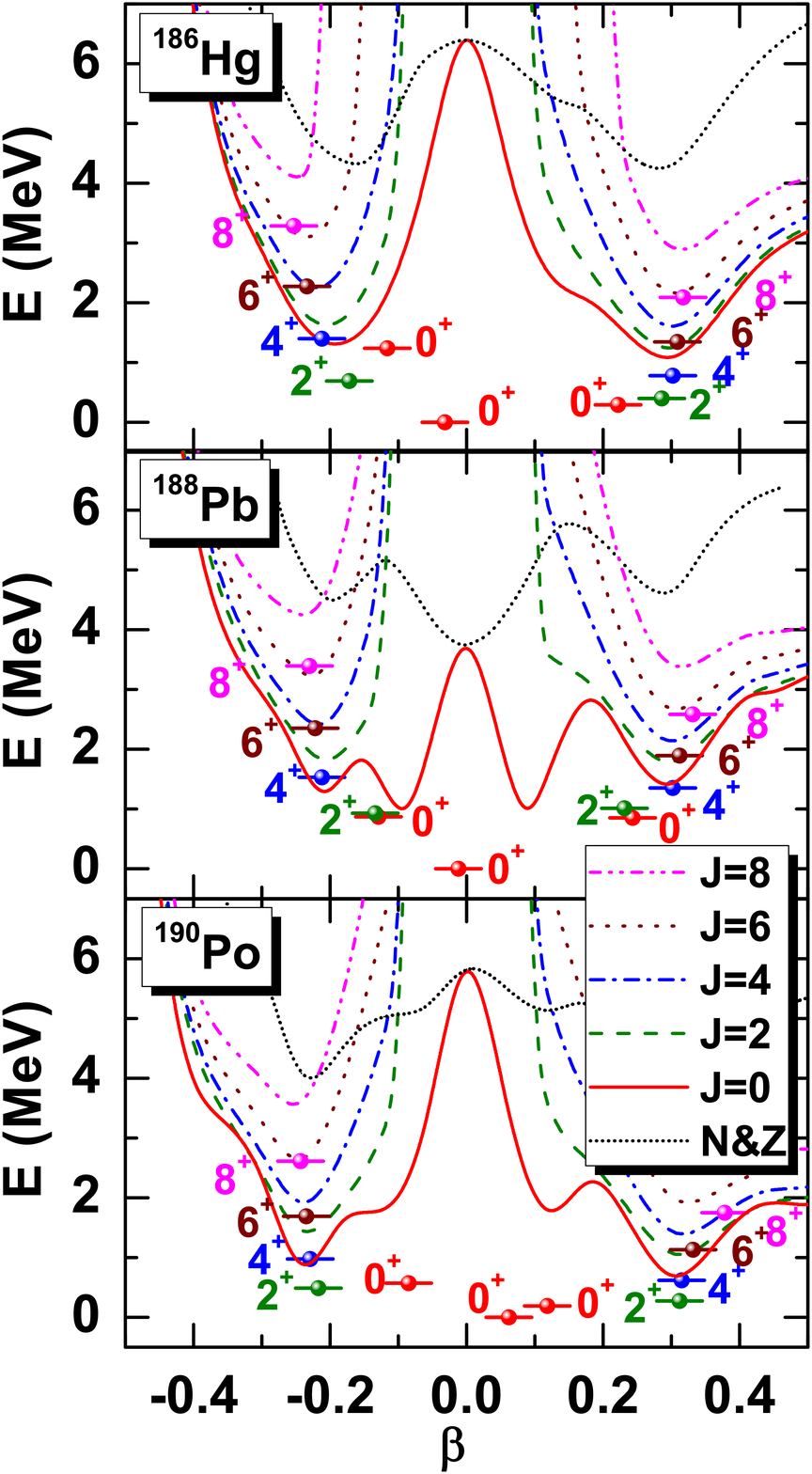}
\caption{ \label{pec:N=106}
(Color online)
Deformation energy curves projected onto particle number (N\&Z),
or additional onto angular momentum ($J=0$, 2, \ldots, 8), as well
as the projected GCM states in the $N=106$ $^{186}$Hg, $^{188}$Pb, and $^{190}$Po isotones.
The projected energy curves are plotted as a function of the intrinsic
deformation $\beta$ of the mean field states. The energies of
projected GCM states are indicated by bullets and horizontal bars
placed at the average deformation $\bar\beta_{Jk}$. All energies
are normalized to the $0^+$ ground state.
}
\end{figure}

\begin{figure*}[t!]
\includegraphics[width=14cm]{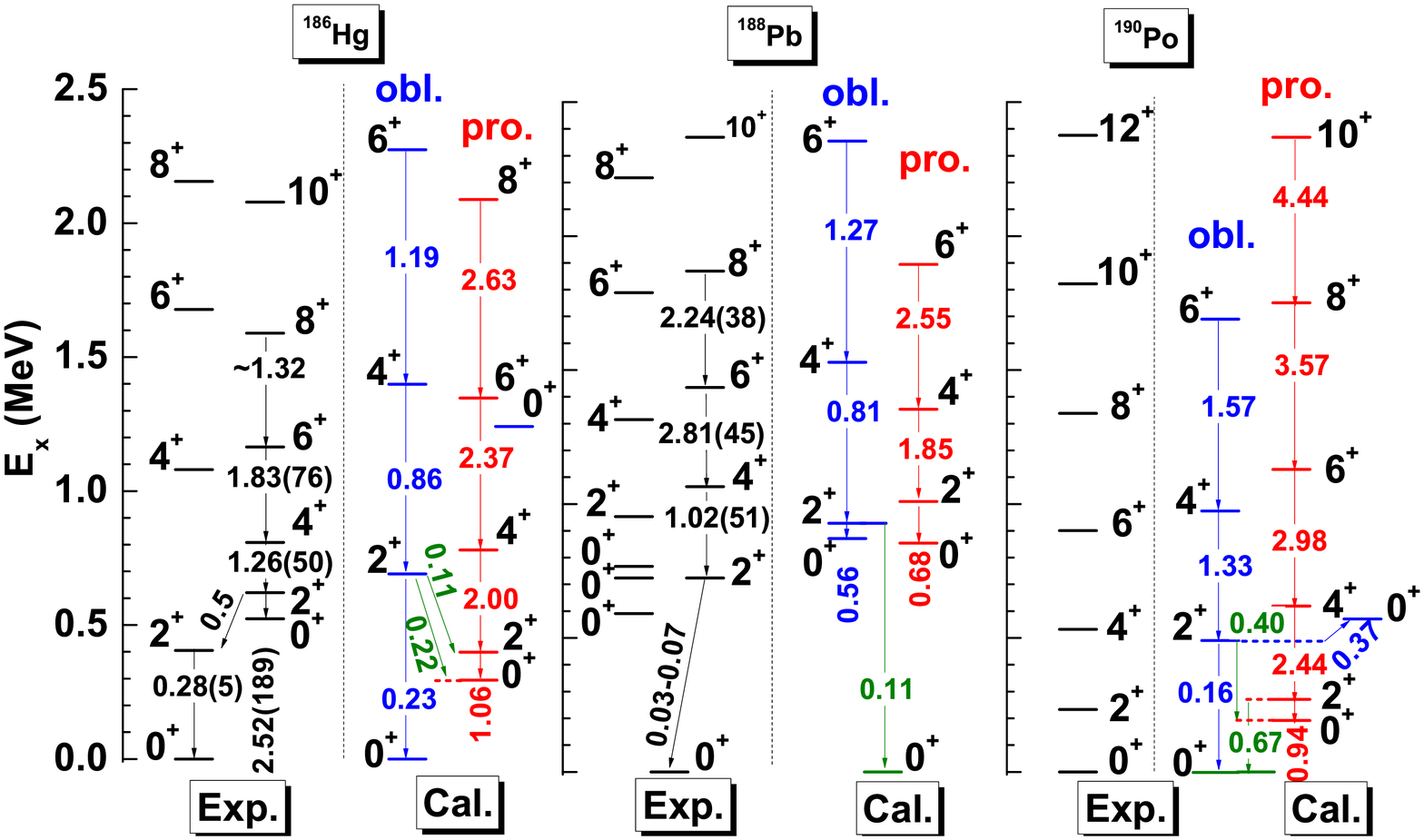}
\caption{\label{spectrum}
(Color online)
Comparison between the calculated and the measured low-lying spectra
for $^{186}$Hg, $^{188}$Pb, and $^{190}$Po. The $B(E2)$ values are
given in $e^2b^2$ units. The experimental data are taken from
Refs.~\cite{Ma93,Vel03PRC,Dewald03,Scheck11,LBL}.
}
\end{figure*}

\subsection{Low-lying states in the $N=106$ isotonic chain}

After the discussion of the $J=0$ states, we will
now turn to the properties of states with finite angular
momentum. Before looking at their systematics, let us first
analyze how these levels group into bands in the case of the
$N=106$ isotones. For this neutron number, our calculations
predict three nearly-degenerate $0^+$ levels for Hg, Pb and Po,
cf.\ Figs.~\ref{wfs1:J0}-\ref{wfs3:J0}, which
makes these isotopes the ideal laboratory for this purpose.

The projected energy curves are plotted in  Fig.~\ref{pec:N=106}
for $^{186}$Hg, $^{188}$Pb, and $^{190}$Po. The energy of the
correlated GCM states are indicated by bars located at their mean
deformation $\bar\beta_{Jk}$. The energy curves projected
only on particle numbers are also provided for comparison. There is no
rotational band on top of the ground state of $^{188}$Pb, which confirms
our interpretation that this level is a correlated spherical state.
However, the mean deformations $\bar\beta_{Jk}$ and energies of the GCM
states, as well as the systematics of $B(E2)$ values and spectroscopic
quadrupole moments to be discussed below, indicate two rotational bands
in this nucleus, one of prolate and the other of oblate deformation.
Similar bands are found for $^{186}$Hg and $^{190}$Po.
For both bands and in all three isotones, the $0^+$ band head is strongly
perturbed.

The energy of the $0^+$ band head is even pushed above that of the $2^+$
state for the oblate band in Hg and Po and quasi-degenerate in Pb. For
larger-$J$ values, the bands, however, become more regular with a steady
increase of the mean deformation $\bar\beta_{Jk}$ as a function of $J$.
This means that deformation follows
closely the minimum of the projected energy curve for the oblate bands,
whereas it is becoming larger for the prolate sequences. In any event,
a third low-lying $0^+$ state that cannot be associated with a rotational
band is also found for $^{186}$Hg and $^{190}$Po.

The theoretical spectra of the three nuclei are compared with the
experimental data in Fig.~\ref{spectrum}. As already discussed in previous
studies~\cite{Bender04}, the calculated spectra are too much expanded.
However, the overall
trends of experiment are rather nicely reproduced. The intraband $E2$
transition strength $B(E2: J+2 \to J)$ increases steadily with angular
momentum, which is consistent with the picture shown in
Fig.~\ref{pec:N=106} that the collective wave functions are mainly
built from either purely prolate or oblate mean-field states as
soon as $J\ge2$. Both the experimental B(E2) values within and
out of bands are reproduced satisfactorily by our calculation.

\begin{figure}[htp]
\includegraphics[width=8cm]{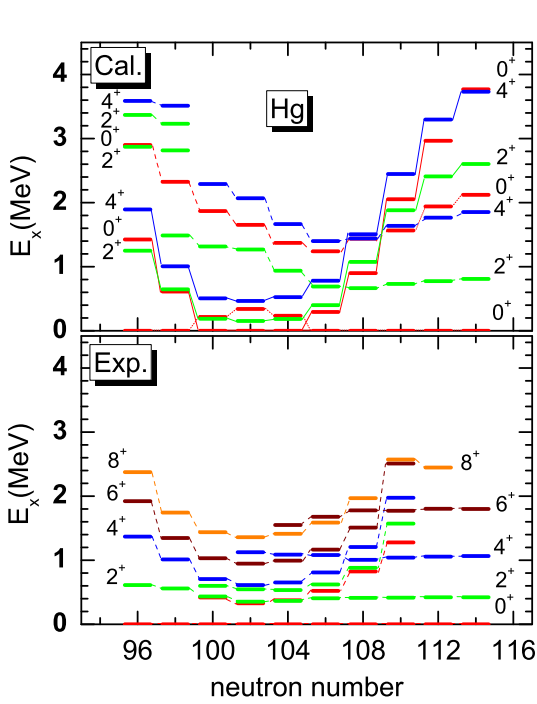}
\caption{\label{Hg:EJ024}
(Color online)
Systematics of calculated (upper panel) and experimental
(lower panel) low-lying $0^+, 2^+$ and $4^+$ states in the Hg
isotopes. The experimental data for the lowest $6^+$ and $8^+$ states
are given as well to indicate the evolution of the band structure.}
\end{figure}

\begin{figure}[htp]
\includegraphics[width=8cm]{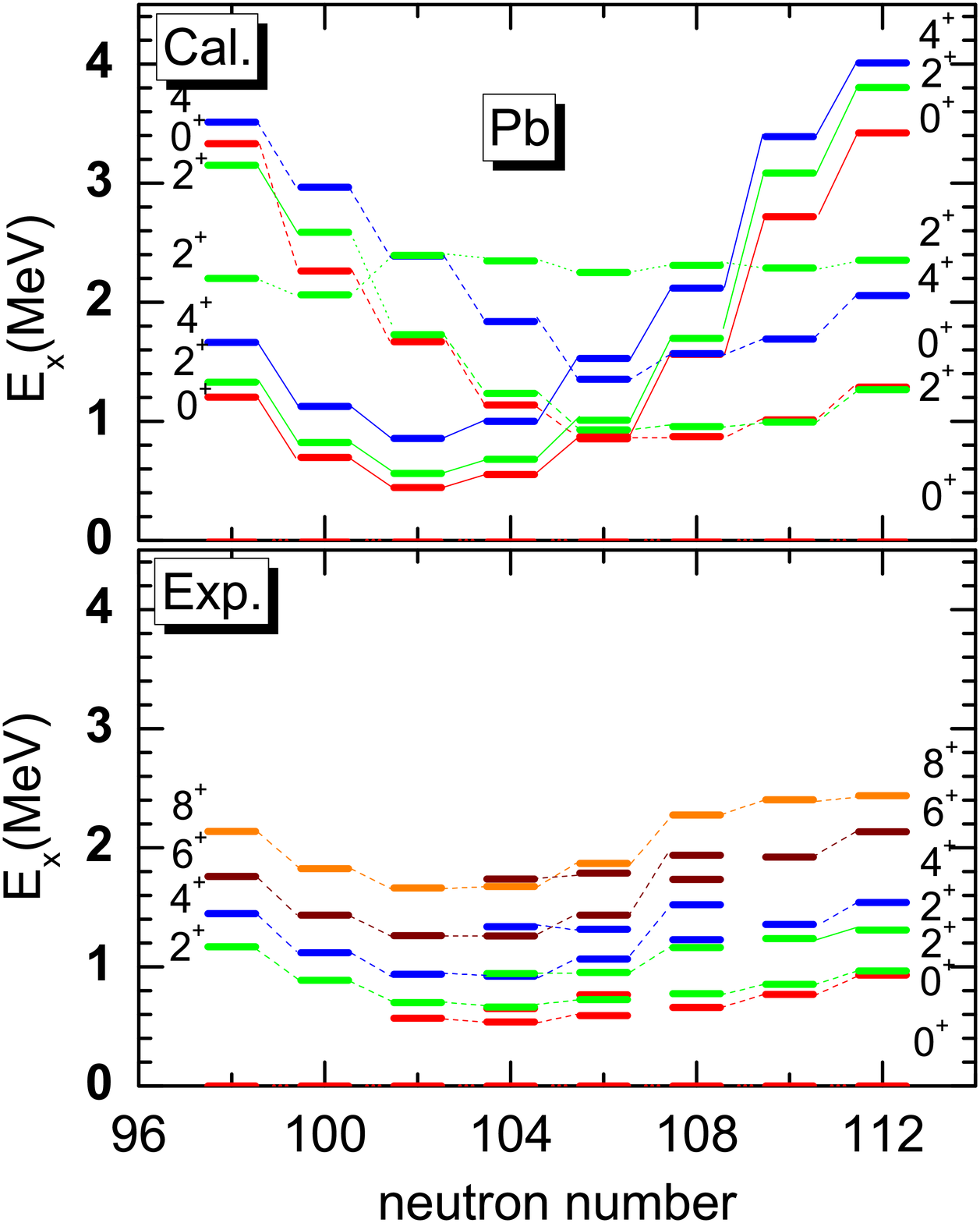}
\caption{\label{Pb:EJ024}
(Color online)
Same as Fig.~\ref{Hg:EJ024}, but for the Pb isotopes.
}
\end{figure}

\begin{figure}[htp]
\includegraphics[width=8cm]{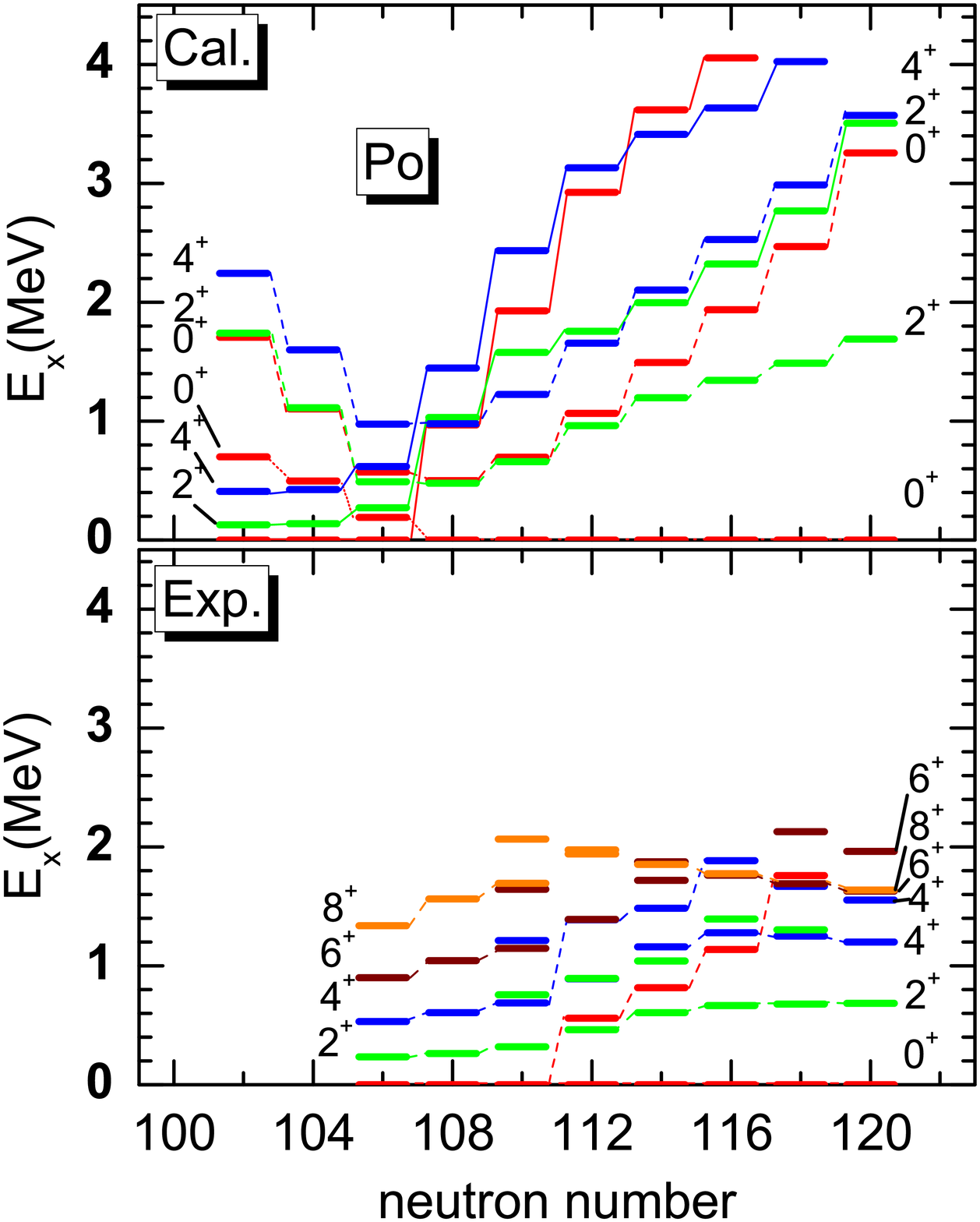}
\caption{\label{Po:EJ024}
(Color online)
Same as Fig.~\ref{Hg:EJ024}, but for the Po isotopes.
}
\end{figure}

\begin{figure}[htp]
\includegraphics[width=8cm]{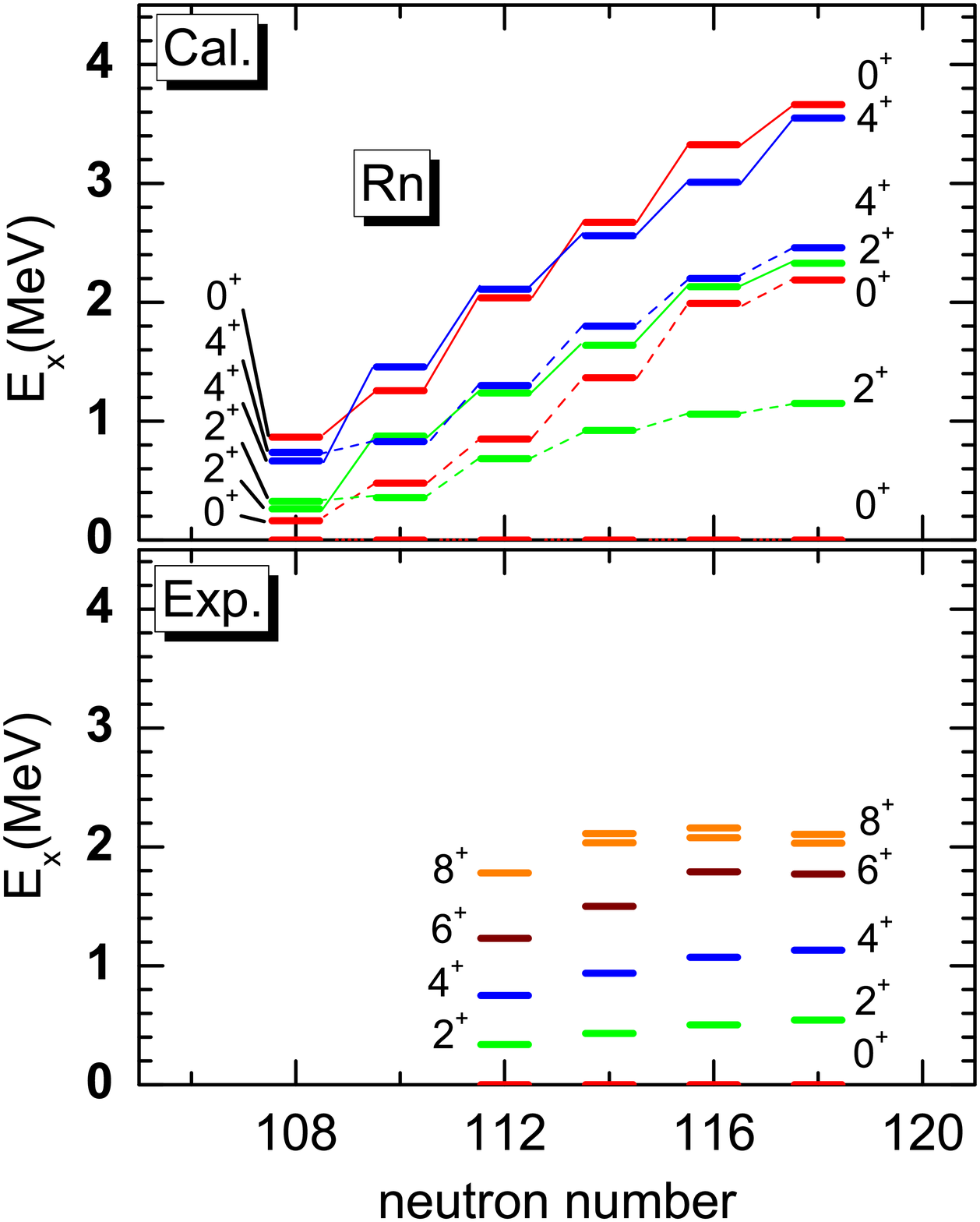}
\caption{\label{Rn:EJ024}
(Color online) Same as Fig.~\ref{Hg:EJ024}, but for the Rn isotopes.
}
\end{figure}

%--------------------------------------------------------------------

\subsection{Systematics of $2^+$ and $4^+$ states}
\label{sect:2+:4+}

The variation as a function of $N$ of the excitation energies of the low-lying
$0^+, 2^+$ and $4^+$ states are presented in
Figs.~\ref{Hg:EJ024}-\ref{Rn:EJ024} for the Hg, Pb, Po and Rn isotopes. The experimental
data are taken from
Refs.~\cite{Dracoulis04,Pakarinen07,Page11,Elseviers11,NNDC}. Lines join
levels with a similar structure in two successive isotopes.

\begin{figure}[t!]
\includegraphics[width=8.2cm]{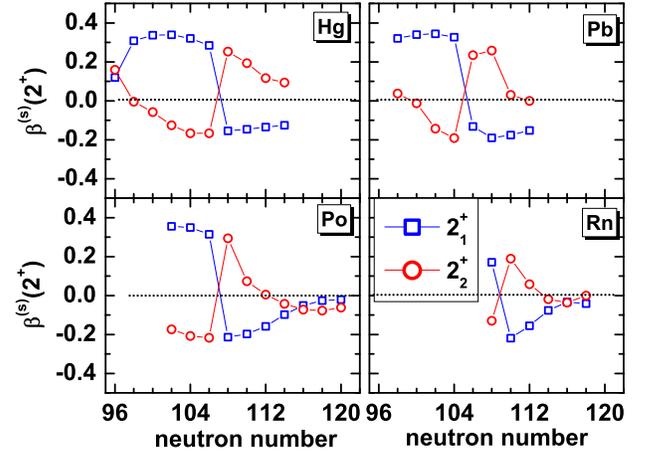}
\caption{\label{beta:sJ2}
The quadrupole deformations $\beta^{(s)}$ (cf.\ Eq.(\ref{beta:s}))
of the two lowest $J=2$ states calculated from the corresponding
spectroscopic quadrupole moments as functions of the neutron number.
}
\end{figure}

\begin{figure}[t!]
\includegraphics[width=8.2cm]{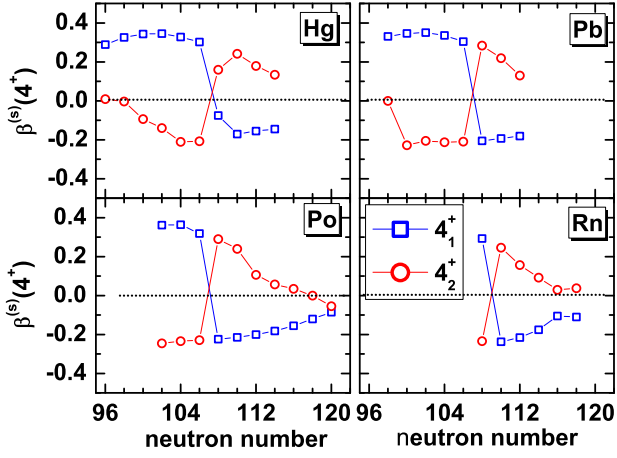}
\caption{\label{beta:sJ4}
Same as Fig~\ref{beta:sJ2}, but for the $J=4$ states.
}
\end{figure}

As already discussed for the \mbox{$N = 106$} isotones, the spectra
are expanded too much.
There can be several sources
to this deficiency: a variational space that is better suited to
optimize the $0^+$ ground state than the excited states, triaxiality
effects, the treatment of pairing correlations or deficiencies of the
energy density functional. We will focus on the evolution of the
relative position of the rotational bands as a function of $N$ (and
sometimes of $Z$), for which the spacing between levels in a band
is in general not crucial. The main findings are as follows:
\begin{itemize}
\item
The experimental trends are reasonably well reproduced by our calculations
for the Hg and the Pb isotopic chains. In Pb, the prolate and the oblate
bands cross between \mbox{$N=108$} and \mbox{$N=106$}, in agreement with
the data. For the Hg isotopes, there is an exchange of structure between
the ground state and the first excited $0^+$ level in such a way that we obtain
a prolate ground state at $N=104$ to $100$, in contradiction with the data.
However, the first two $0^+$ states remain close in energy for these three
isotopes.

\item
In Hg and Pb, the mixing between the $0^+$ levels is usually too large
in the calculations, distorting too much the rotational character of bands
at low spin. Experimentally, the energy of the first $2^+$ level is nearly
constant over the entire isotopic Hg chain, whereas its variation is much
larger in our calculations. The energy difference between the $2^+$ and
$4^+$ levels varies in a manner more similar to the data than the excitation
energy of the $2^+$ level.

\item
In the Hg and Pb isotopic chains, the calculated excitation energies of
the $2^+_1$ and $4^+_1$ levels reach their lowest value at \mbox{$N=102$}.
This result agrees with the experimental data for the Hg isotopes.
For the Pb isotopes, the lowest excitation energy for these levels
is obtained in the experimental data at $N=104$, but the calculated
results at $N=102$ and $104$ are only marginally different.

\item
There is a clear disagreement between calculations and experimental
data for the heavy Po isotopes. The overestimation of energies is close to
a factor of two and much larger than for other nuclei. In particular, the
near-constant value of the experimental $2^+$ level energy between \mbox{$N=120$}
and $112$ is not reproduced at all. Being close to the \mbox{$N=126$} shell
closure, these isotopes are probably not collective and would require to
include $n$p-$m$h excitations that are not generated by a constraint on
deformation. The calculated spectra are more realistic for the isotopes
with \mbox{$116 \leq N \leq 106$}.

\item
The disagreement with the data is not as strong for the Rn isotopes. In
particular, the energy difference between the $2^+$ and $4^+$ levels is closer
to the experimental value.

\end{itemize}
The quadrupole deformation parameters determined from the
spectroscopic quadrupole moment, $\beta^{(s)}$ (cf.\ Eq.(\ref{beta:s})),
are plotted as a function of the neutron number in Figs.~\ref{beta:sJ2}
and \ref{beta:sJ4} for the lowest two $2^+$ and $4^+$ states, respectively.
This parameter, which is directly related to an observable, permits a much
less ambiguous assignment of a prolate or an oblate character to the
levels than the mean deformation $\bar\beta_{Jk}$ of the wave functions.
A crossing between oblate and prolate configurations takes place at
\mbox{$N=106$} for the Hg, Pb and Po isotopes, at \mbox{$N=108$} for Rn.
Again, this result is not confirmed by the data for the Hg isotopes. For
the heaviest Po and Rn isotopes,  $\beta^{(s)}$ is very small. This is a
sign that the $2^+$ and $4^+$ are probably not based on a deformed state,
but would be better described by $n$p-$m$h excitations of the kind already
mentioned above. This could also be at the origin of the much too large
excitation energies that we have obtained for these $2^+$ and $4^+$ states
(see Fig.~\ref{Po:EJ024} and~\ref{Rn:EJ024}); our purely collective GCM
basis not being well-suited to describe these states.

%--------------------------------------------------------------------

\begin{figure}[]
\includegraphics[width=9cm]{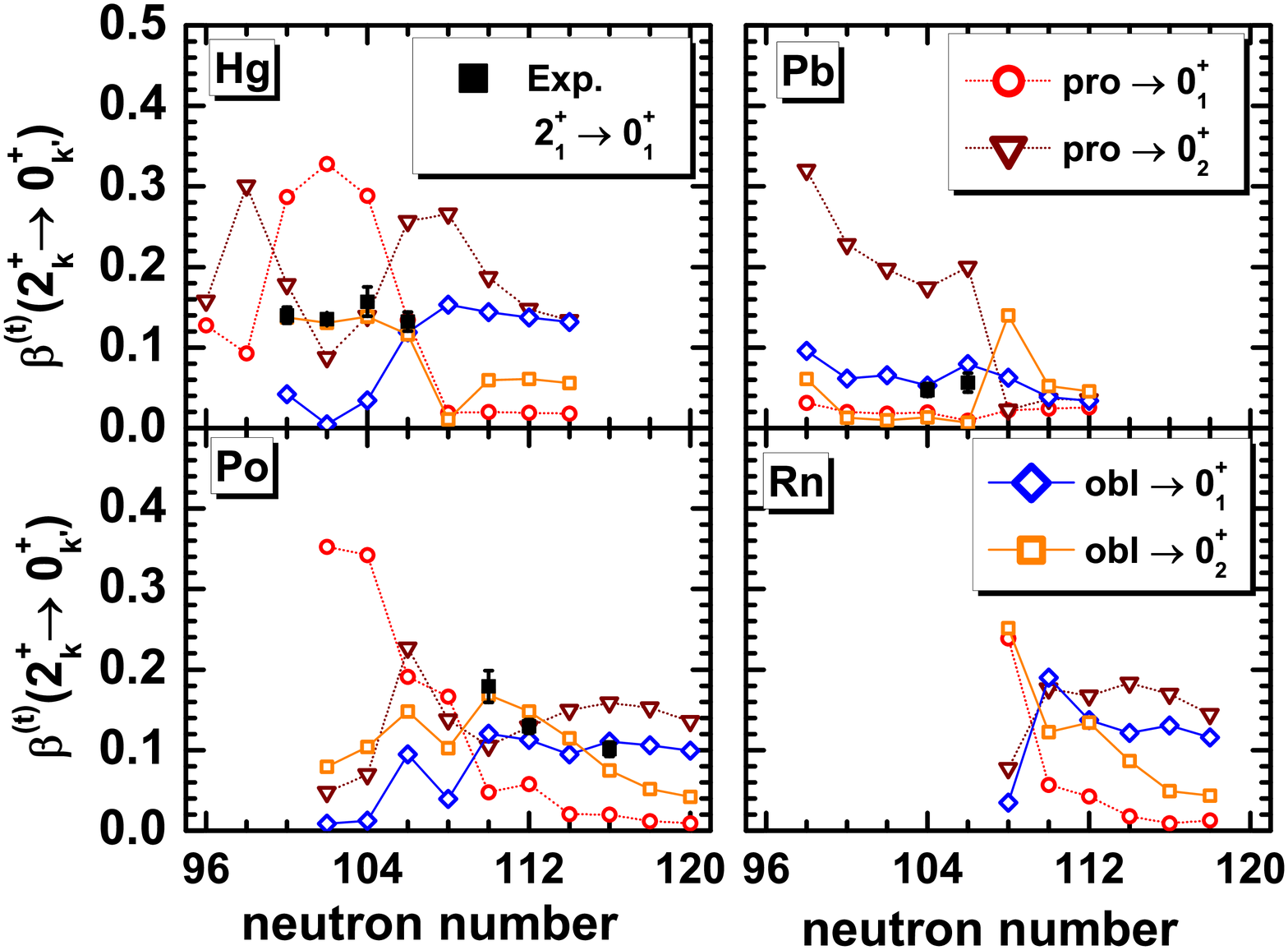}
\caption{\label{BE2}
(Color online) Systematics of the deformations $\beta^{(t)}$ derived
from the reduced $E2$ strengths for the transition from the predominant
prolate and oblate $2^+$ states to the first two $0^+$ states as a
function of neutron number. The data for transition from the $2^+_1$
state to the ground state are taken from
Refs.~\cite{Rud73,Ellegaard73,Dewald03,Grahn06,Grahn09,Kesteloot10}.
}
\end{figure}

The quadrupole deformations $\beta^{(t)}(2^+_k \rightarrow 0^+_{k'})$, cf.\ Eq.(\ref{beta:t}), corresponding to the
electric quadrupole transition strengths $B(E2:2^+_k \rightarrow 0^+_{k'})$ are plotted in
Figure~\ref{BE2} for the transitions from the two first $2^+$ to the
 $0^+$ ground state and the $0^+_2$ level, together with the available experimental data.
The assignment of oblate and prolate labels to the $2^+$ states are based on spectroscopic moments and are, thus, non
ambiguous. The main findings are as follows:
\begin{itemize}

\item
The systematics of $\beta^{(t)}$ values for transitions between oblate or prolate states
are quite similar in the four isotopic chains.

\item
For $^{180,182,184}$Hg, the experimental $\beta^{(t)}(2^+_1 \to 0^+_1)$
values are in good agreement with the calculated ones for the transition
from the oblate $2^+$ level to the oblate $0^+_2$ state.
It shows also that theoretical transition probabilities can be
accurate in spite of the fact that the energies and the order of levels
are not reproduced.

\item
For the Pb isotopes with $N=104$ and $106$, the experimental
$\beta^{(t)}(2^+_1 \to 0^+_1)$ value is in good agreement with the
calculated transition from the oblate $2^+$ state to the spherical
$0^+_1$ state. It suggests that, with decreasing neutron number, the
$2^+_1$ state keeps predominantly an oblate character down to $N=104$,
beyond which the prolate $2^+$ state becomes the yrast state
in the calculations.

\item
For the $N=116$ Po isotope, the experimental $\beta^{(t)}(2^+_1 \to 0^+_1)$
value is reproduced quite well by the calculation. It corresponds to a
transition from the oblate $2^+_1$ level to the spherical $0^+_1$ state.
As the neutron number decreases to \mbox{$N=112$} and $110$, the transitions
calculated to the oblate $0^+_2$ state are  in better agreement with
the experimental $\beta^{(t)}(2^+_1 \to 0^+_1)$ values than the transitions
to the spherical $0^+_1$ state. This indicates an onset of oblate deformation
already at \mbox{$N=112$}, which is consistent with the observation made
from the experimental isotope shifts, c.f.\ Fig.~\ref{isotope-shifts}.

\item
There are no experimental data for the $B(E2)$ values in the
Rn isotopes discussed here.

\end{itemize}

\subsection{Kinetic moment of inertia along the yrast band}

\begin{figure}[]
\includegraphics[width=8.2cm]{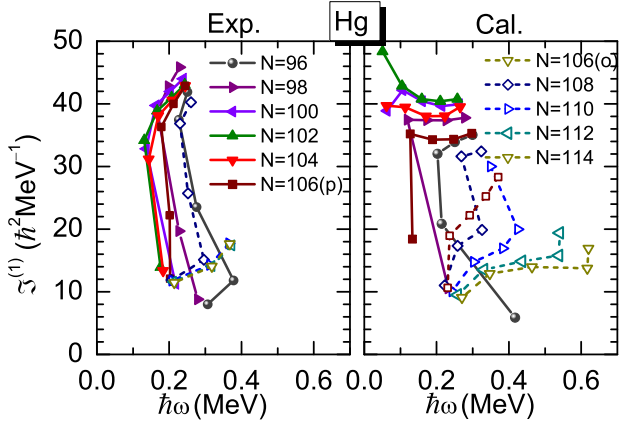}
\caption{
Kinematic moment of inertia $\Im^{(1)}(J)$ for Hg isotopes,
as a function of the frequency $\omega(J\to J-2)$ for transitions along
the yrast line. For the calculated values on the right, $\hbar\omega$ has
been scaled by a factor 1.50. Data on the left panel are  taken from
Ref.~\cite{NNDC}. }
\label{Hg:MOI}
\end{figure}

\begin{figure}[]
\includegraphics[width=8.2cm]{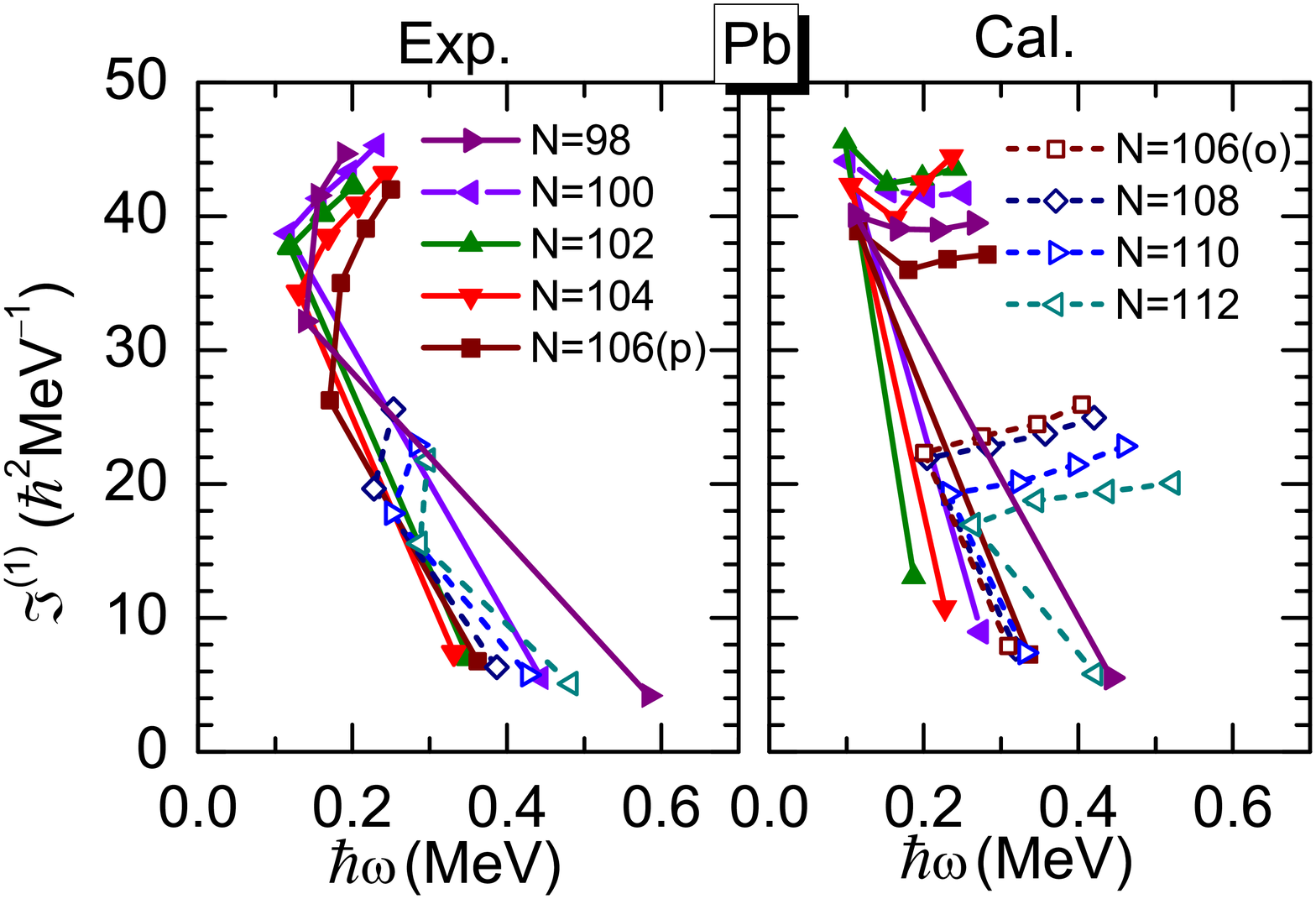}
\caption{Same as Fig~\ref{Hg:MOI}, but for the Pb isotopes.}
\label{Pb:MOI}
\end{figure}

\begin{figure}[]
\includegraphics[width=8.2cm]{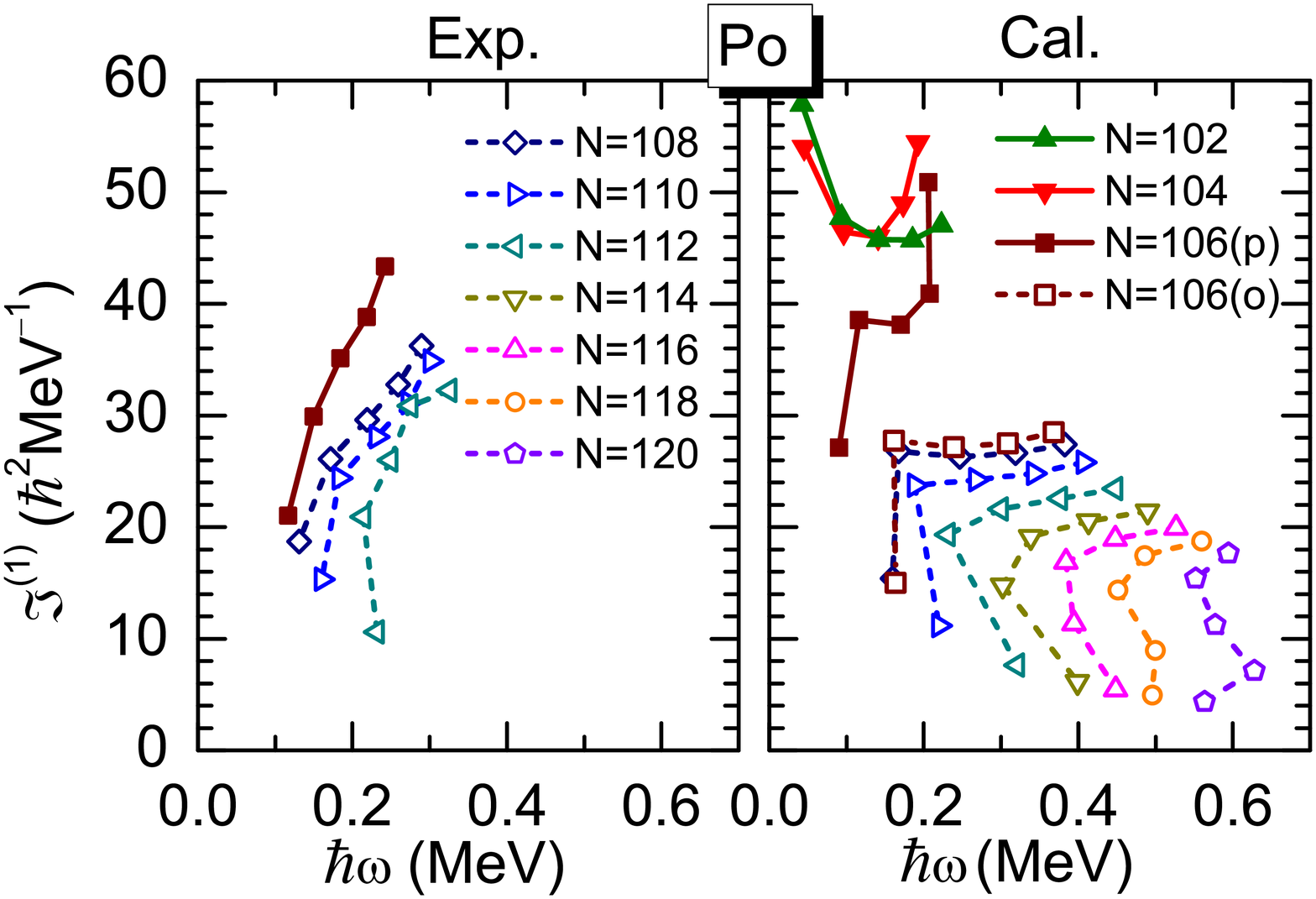}
\caption{Same as Fig~\ref{Hg:MOI}, but for the Po isotopes.}
\label{Po:MOI}
\end{figure}

\begin{figure}[]
\includegraphics[width=8.2cm]{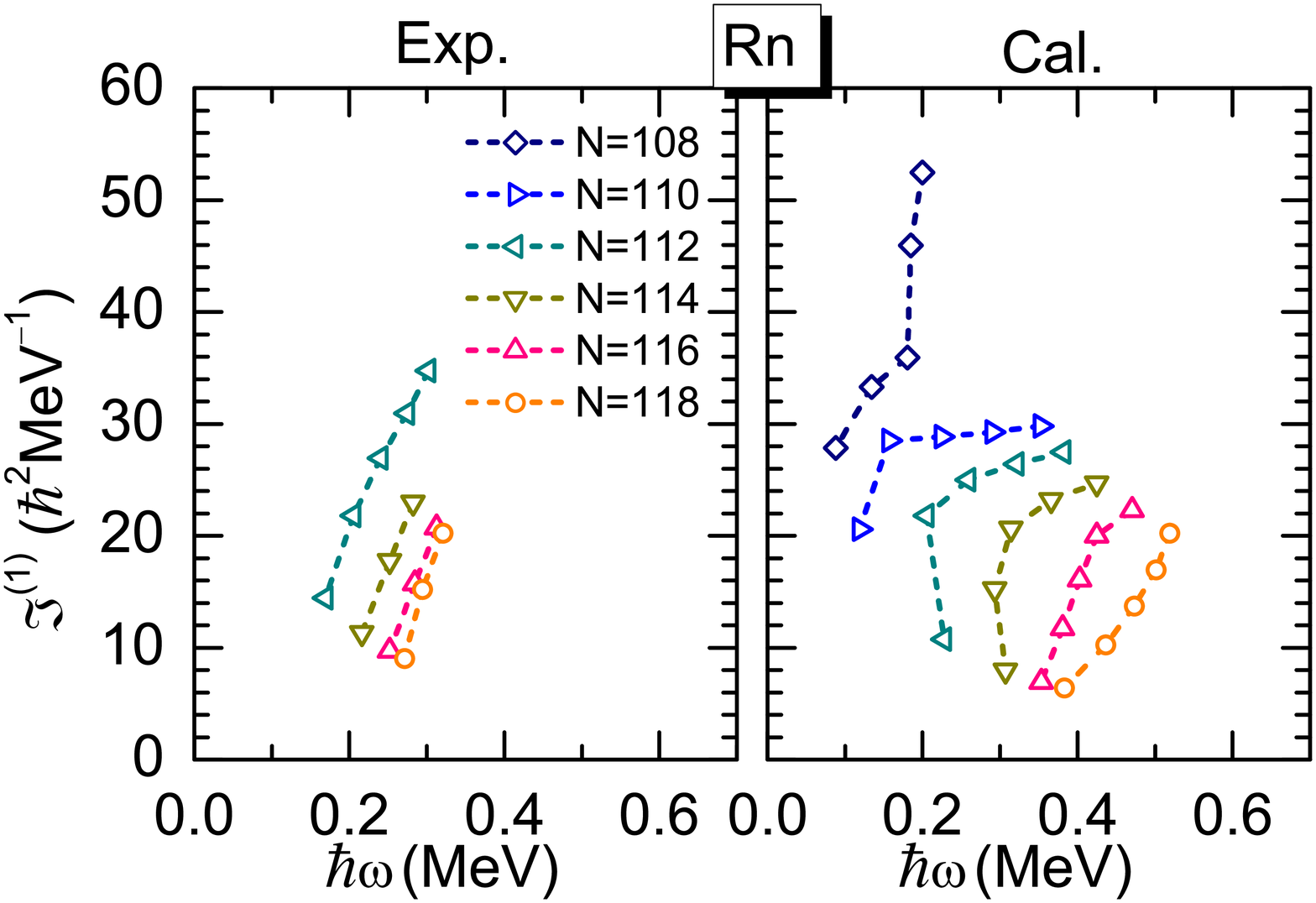}
\caption{Same as Fig~\ref{Hg:MOI}, but for the Rn isotopes.}
\label{Rn:MOI}
\end{figure}

The variation of the moment of inertia along a band can be an indicator
of its nature. There are well
established generic properties of the
moments of inertia such
as their increase with deformation, or that for
the same value of $| \beta |$ the moment of inertia is larger for
prolate deformations than for oblate ones. The kinetic moment of inertia
$\Im^{(1)}$ is defined as

\begin{equation}
\label{MoI}
\Im^{(1)}(J)
=  \frac{\hbar \, \sqrt{J(J+1)}}
        {\omega(J\to J-2)}
\end{equation}
and the frequency $\omega$,
\begin{equation}
\label{frequency}
\hbar\omega (J\to J-2)
= \tfrac{1}{2} \; [E_x(J)- E_x(J-2)] \, .
\end{equation}
The kinematic moments of inertia for transitions along the yrast
line are plotted in Fig.~\ref{Hg:MOI}-\ref{Rn:MOI} for transitions from \mbox{$J=2$}, 4, \ldots, 10
to $J-2$ in Hg, Pb, Po and Rn isotopes,
respectively. As discussed above, the calculations underestimate
the moment of inertia, cf.\ Fig.~\ref{spectrum}.
To make the comparison  with the experimental data easier, we have
divided the calculated rotational frequency by a factor 1.50.

There are several common features in the four isotopic
chains:
\begin{itemize}

\item
For a static rotor, the kinematic moment of inertia is constant. The only
nuclei in our sample which came close to this idealized case are the Hg
isotopes with \mbox{$N = 112$} and 114, at least up to \mbox{$J=8$} in both
calculation and experiment. For all other nuclei the pattern is more
complex.

\item
The kinematic moment of inertia $\Im^{(1)}$ for \mbox{$J=2$} is significantly
different from those corresponding to transitions from
higher spins. This can be attributed to the ground state configuration
that in most cases is dominated neither by an oblate nor by a prolate
configuration. In the Pb isotopes, the ground state is a correlated spherical
state, with a very different structure from the first $2^+$ state. In any
case, the ground state cannot be considered as the band head of a
rotational band.

\item
The kinematic moments of inertia
$\Im^{(1)}$  of states with spin \mbox{$J \ge 4$} can be divided into
two groups: the isotopes with \mbox{$N \le 106$} have a prolate
configuration and a larger $\Im^{(1)}$, whereas the isotopes with
\mbox{$N > 108$} are oblate and have a smaller $\Im^{(1)}$. This
feature is present in both the calculations and the data. This confirms
again that prolate configurations dominate the yrast states below
\mbox{$N=108$}.

\item
For many nuclei, the calculated $\Im^{(1)}$ moments take almost constant
values for the highest angular momenta studied here. Such behavior,
however, is almost never seen in the data. This discrepancy can be
expected to have the same origins as the overall expansion of the
spectra, cf.\ Sect.~\ref{sect:2+:4+}.

\end{itemize}

\section{Summary and outlook}
\label{Sec.IV}

We have shown that a detailed comparison between the results
obtained by a beyond mean-field method and experimental spectroscopic data
can be performed for a large number of nuclei. This comparison can be carried out
without ambiguities when restoring quantum numbers and thereby
selection rules for transitions. The region around the neutron-deficient
Pb isotopes is of particular interest for such studies because it
presents a very rich variety of phenomena, and the structure
of these nuclei varies rapidly as a function of the neutron number.

The results of our model are in qualitative agreement with the
experimental data. We obtain the coexistence of several low-lying $0^+$
levels showing mixing between oblate, prolate and spherical configurations.
The order of these levels is sometimes different from that
deduced from the analysis of the data for $\alpha$-decay hindrance,
radii, moments of inertia and electromagnetic transition densities.
However, the theoretical $B(E2)$ values within a rotational band of
given prolate or oblate nature are usually described well even though
the relative position of the band heads is not correct.

We find that labeling states as oblate or prolate might not always be
adequate, especially for nuclei in this mass region. In particular,
the $0^+$ and $2^+$ states often result from a mixing between
symmetry-restored mean-field wave functions corresponding to shapes spread
over a wide range of deformations. The role of the mixing diminishes
with increasing angular momentum, such that the spectra have more
apparent rotational or vibrational character at high spin.

Our study has been limited to the mixing of particle-number and
angular-momentum restored time-reversal invariant axial configurations.
Tools to extend such calculations to triaxial shapes have been set-up
recently~\cite{Bender2008,Yao10,Rodriguez10,Yao11,Rod11a,Bau12a}.
However, the presently available computational resources do not yet
allow for their systematical application to very heavy nuclei. We
have seen that triaxial configurations might play a role for some of
the lightest Hg and some of the heaviest Po and Rn isotopes studied here.
The absence of this degree of freedom is one of
the reasons for the systematically too expanded theoretical excitation
spectra. To further improve the description of excited states,
a second extension of the variational space to be considered in the future
is the projection and mixing of time-reversal invariance-breaking mean-field
states. On the one hand, starting from cranked HFB states that are
optimized for finite angular momenta will incorporate the rotational
alignment of single-particle states and the weakening of pairing
correlations at high spin that comes with it. And on the other hand,
the use of so-called blocked HFB states of the broken-pair type will provide
a natural starting point to describe the non-collective states that can
be suspected to dominate the low-lying excitation spectra of the heaviest
Po and Rn isotopes discussed here. Developments in both directions are
underway, but, again, first applications will
be limited to light nuclei for computational reasons.

The energy density functionals currently used do not permit yet to reach
an accurate quantitative description of spectroscopic data. However, we
recall that
the energy density functional used here has quite a simple form and that its parameters have
been adjusted on the mean-field level to bulk properties of nuclear matter
and magic nuclei. Hence, the overall good qualitative description of the
rapidly evolving collective states of the nuclei studied here should be
viewed as a success. This can be partially attributed to the generic features
of the single-particle level schemes that remain an efficient tool to
qualitatively analyze also results from beyond-mean-field calculations,
as has been shown by our discussion. In addition to
the well-known correlation between the opening and closing of
gaps around the Fermi energy in the Nilsson diagram and the appearance
of the various minima in the deformation energy surfaces, we have also seen
that the spread of the correlated collective wave functions over different
deformations is correlated with the deformations at which the intruder levels
cross the single-particle levels occupied in the spherical configuration.
In general, we find that the mixing of states with a different number of
occupied intruder levels is disfavored, which is consistent with the usual
classification of shape coexisting states in the interacting shell model.
A better quantitative description of fine details of shape coexistence,
such as the relative energy between oblate and prolate band heads in
the Hg isotopes below \mbox{$N = 108$}, will require improved single-particle
energies, cf.\ also the discussion in Ref.~\cite{Bender06}.
First explorative studies in this direction indicate, however, that it is
unlikely to obtain a significantly better description of single-particle
energies by a refit within the current standard form of the energy density functionals
\cite{Lesinski07,Kortelainen08}, and that extensions are needed. Work in
this direction is also underway.

\begin{acknowledgments}

We thank R. Janssens and P. Van Duppen for useful comments on this manuscript.
This research was supported in parts by the PAI-P6-23 of the Belgian Office for Scientific Policy, the National Science Foundation of China under Grants No.~11105111 and
No.~10947013, the Fundamental Research Funds for the Central Universities (XDJK2010B007), the European Union's Seventh Framework Programme ENSAR under grant agreement n262010,
by the French Agence Nationale de la Recherche under Grant No.~ANR
2010 BLANC 0407 "NESQ", and by the CNRS/IN2P3 through the PICS No.~5994.
A part of the computations was performed using HPC resources from
GENCI-IDRIS (Grants No.\ 2007-050707, 2008-050707, and 2009-050707).

\end{acknowledgments}

\end{document}